\documentclass[10pt,letterpaper]{article}

\usepackage{changepage}
\usepackage[utf8]{inputenc}
\usepackage{textcomp,marvosym}
\usepackage{fixltx2e}
\usepackage{amsmath,amssymb}
\usepackage{cite}
\usepackage{nameref,hyperref}
\usepackage[right]{lineno}
\usepackage{microtype}
\DisableLigatures[f]{encoding = *, family = * }
\usepackage{rotating}
\usepackage{color}
\usepackage{graphicx}
\usepackage{graphics}

\usepackage{tikz}
\usetikzlibrary{positioning}

\makeatletter
\renewcommand{\@biblabel}[1]{\quad#1.}
\makeatother

\date{}

\usepackage{color}

\begin{document}
% \vspace*{0.35in}

% Title must be 150 characters or less
\begin{flushleft}
{\Large
%\textbf\newline{Analysing Rumour Diffusion, Support and Denial in Social Media}
\textbf\newline{Analysing How People Orient to and Spread Rumours in Social Media by Looking at Conversational Threads}
}
\newline
% Insert Author names, affiliations and corresponding author email.
\\
Arkaitz Zubiaga\textsuperscript{1}*,
Maria Liakata\textsuperscript{1},
Rob Procter\textsuperscript{1},
Geraldine Wong Sak Hoi\textsuperscript{2},
Peter Tolmie\textsuperscript{1}
\\
\bf{1} University of Warwick, Gibbet Hill Road, CV4 7AL Coventry, UK
\\
\bf{2} swissinfo.ch, Bern, Switzerland
\\
%\textcurrency University of Warwick, Coventry, United Kingdom

* a.zubiaga@warwick.ac.uk

% Insert additional author notes using the symbols described below. Insert symbol callouts after author names as necessary.
% 
% Remove or comment out the author notes below if they aren't used.
%
% Primary Equal Contribution Note
% \Yinyang These authors contributed equally to this work.

% Additional Equal Contribution Note
% \ddag These authors also contributed equally to this work.

% \textcurrency b Insert current address of second author with an address update
% \textcurrency c Insert current address of third author with an address update

% \dag Deceased

% Group/Consortium Author Note
% \textpilcrow Insert Collaborative Author line here

\end{flushleft}
\section*{Abstract}
As breaking news unfolds people increasingly rely on social media to stay abreast of the latest updates. The use of social media in such situations comes with the caveat that new information being released piecemeal may encourage rumours, many of which remain unverified long after their point of release. Little is known, however, about the dynamics of the life cycle of a social media rumour. In this paper we present a methodology that has enabled us to collect, identify and annotate a dataset of 330 rumour threads (4,842 tweets) associated with 9 newsworthy events. We analyse this dataset to understand how users spread, support, or deny rumours that are later proven true or false, by distinguishing two levels of status in a rumour life cycle i.e., before and after its veracity status is resolved. The identification of rumours associated with each event, as well as the tweet that resolved each rumour as true or false, was performed by a team of journalists who tracked the events in real time. Our study shows that rumours that are ultimately proven true tend to be resolved faster than those that turn out to be false. Whilst one can readily see users denying rumours once they have been debunked, users appear to be less capable of distinguishing true from false rumours when their veracity remains in question. In fact, we show that the prevalent tendency for users is to support every unverified rumour. We also analyse the role of different types of users, finding that highly reputable users such as news organisations endeavour to post well-grounded statements, which appear to be certain and accompanied by evidence. Nevertheless, these often prove to be unverified pieces of information that give rise to false rumours. Our study reinforces the need for developing robust machine learning techniques that can provide assistance in real time for assessing the veracity of rumours. The findings of our study provide useful insights for achieving this aim.

% \linenumbers

\section*{Introduction}

Social media have increasingly gained popularity in recent years, enabling people not only to keep in touch with family and friends, but also to stay abreast of ongoing events and breaking news as they unfold. The potential for spreading information quickly through a large community of users is one of the most valuable characteristics of social media platforms. Social media, being open to everyone, enable not only news organisations and journalists to post news stories, but also ordinary citizens to report from their own perspectives and experiences. This broadens the scope and diversity of information that one can get from social media and sometimes may even lead to stories breaking before they appear in mainstream media outlets \cite{kwak2010twitter}. While this often leads to having access to more comprehensive information, it also comes with caveats, one of which is the need to sift through the different information sources to assess their accuracy \cite{silverman2013verification}.

The spread of misinformation is especially important in the context of breaking news, where new pieces of information are released piecemeal, often starting off as unverified information in the form of a rumour. These rumours then spread to large numbers of users, influencing perception and understanding of events, despite being unverified. Social media rumours that are later proven false can have harmful consequences both for individuals and for society \cite{zubiaga2014tweet}. For instance, a rumour in 2013 about the White House having been bombed, injuring Barack Obama, which was tweeted from AP's Twitter account by hackers, spooked stock markets in the US\footnote{http://www.bbc.co.uk/news/world-us-canada-21508660}. A major event that was similarly riddled with consequential rumours was Hurricane Sandy, which hit the East Coast of the US in 2012. Part of the city of New York suffered from power outages and many people had to rely on the Internet accessed through their mobile phones for information. To prevent major incidents, the US Federal Emergency Management Agency had to set up a web page specifically for rumour control\footnote{https://twitter.com/fema/status/264800761119113216}.

While rumours in social media are a concern, little work has been done so far to understand how they propagate. In this work we aim to help rectify this by examining in some detail rumours generated on Twitter within the context of nine different newsworthy events. First, we introduce our methodology to collect, identify, annotate and enable the analysis of social rumours. Then, we look at how rumours are spawned, and later confirmed or debunked; our study looks at conversations around rumours in social media, exploring how social media users respond to rumours both before and after the veracity of a rumour is resolved. Our study provides insight into rumour diffusion, support and denial in social media, helping both those who gather news from social media in determining accuracy of information and the development of machine learning systems that can provide assistance in real-time for assessing the veracity of rumours \cite{derczynski2014pheme}. 

\subsection*{Background}

One of the main challenges when studying rumours is to come up with a sound definition of the concept. In contrast to recent research that considers a rumour as constituting information that is ultimately deemed false, here we emphasise our adherence to the widely accepted definition of rumours as ``unverified and instrumentally relevant information statements in circulation'' \cite{difonzo2007rumor}. This unverified information may turn out to be true, or partly or entirely false. Adapting the existing definition to our context of breaking news stories, we define a rumour as a ``\textit{circulating story of questionable veracity, which is apparently credible but hard to verify, and produces sufficient skepticism and/or anxiety so as to motivate finding out the actual truth}'' \cite{zubiaga2015towards}.

Rumours and related phenomena have now been studied from many different perspectives \cite{donovan2007idle} ranging from psychological studies \cite{rosnow2005rumor} to computational analyses \cite{qazvinian2011rumor}. Traditionally, it has been very difficult to study people's reactions to rumours, given that this would involve real-time collection of reaction as rumours unfold. To overcome this obstacle, Allport undertook early investigations \cite{allport1946analysis,allport1947psychology} in the context of wartime rumours. He posited the importance of studying rumours, emphasising that ``newsworthy events are likely to breed rumors'' and that ``the amount of rumor in circulation will vary with the importance of the subject to the individuals involved times the ambiguity of the evidence pertaining to the topic at issue''. This led him to set forth a motivational question which is yet to be answered: ``Can rumors be scientifically understood and controlled?'' \cite{allport1946analysis}. His 1947 experiment \cite{allport1947psychology} reveals an interesting fact about rumour circulation and belief. He looked at how US President Franklin D. Roosevelt allayed rumours about losses at the 1941 Pearl Harbor bombing. The study showed that before the President made his address, 69\% of a group of undergraduate students believed that losses at Pearl Harbor were greater than officially stated; but five days later, the President having spoken in the meantime, only 46\% of an equivalent group of students believed this statement to be true. This study revealed the importance of an official announcement by a reputable person in shaping society's perception of the accuracy of a rumour.

Early research focused on different objectives. Some work has looked at the factors that determine the diffusion of a rumour, including, for instance, the influence of the believability of a rumour on its subsequent circulation, where believability refers to the extent to which a rumour is likely to be perceived as truthful. Early research by Prasad \cite{prasad1935psychology} and Sinha \cite{sinha1952behaviour} posited that believability was not a factor affecting rumour mongering in the context of natural disasters. More recently, however, Jaeger et~al. \cite{jaeger1980hears} found that rumours were passed on more frequently when the believability level was high. Moreover, Jaeger et~al. \cite{jaeger1980hears} and Scanlon \cite{scanlon1977post} found the importance of a rumour as perceived by recipients to be a factor that determines whether or not it is spread, the least important rumours being spread more. Others have attempted to categorise different types of rumours. Knapp \cite{knapp1944psychology} introduced a taxonomy of three types of rumours: (1) `pipe-dream' rumours, as rumours that lead to wishful thinking, (2) `bogy' rumours, as those that increase anxiety or fear, and (3) `wedge-driving' rumours, as those that generate hatred.

% \cite{rosnow1988factors}
% \begin{itemize}
%  \item ``the results of some recent studies have implied that subjects were, if anything, more inclined to spread rumors that they did not consider to be important (Jaeger et al., 1980; Scanlon, 1977)''
%  \item Jaeger et al., 1980 found that rumors were more frequently repeated in the high than in the low belief condition.
%  \item However, Prasad (1935) and Sinha (1952) concluded that credulity was not a factor affecting rumor mongering in the context of natural disasters.
% \end{itemize}

% \cite{knapp1944psychology}
% \begin{itemize}
%  \item Patterns worth noting in rumours:
%  \begin{itemize}
%   \item Exhibitionism. This is undoubtedly one of the prime motives. The insecure person may often resort to rumor-mongering as a means of heightening his personal status with his associates. By disseminating rumors he conveys the impression that he has access to information and knowledge not available to others.
%   \item Bestowing a favor.
%   \item Reassurance and emotional support. Here the rumor is told primarily for the purpose of sharing one's own anxiety with another.
%   \item Aggression.
%   \item Projection of subjective conflicts.
%   \item ``Partial facts, or facts which are too fragmentary to permit interpretation can spawn rumors.''
%  \end{itemize}
% \end{itemize}

The widespread adoption of the Internet gave rise to a new phase in the study of rumour in naturalistic settings \cite{bordia1996studying} and has taken on particular importance with the advent of social media, which not only provides powerful new tools for sharing information but also facilitates data collection from large numbers of participants. For instance, Takayasu et~al. \cite{takayasu2015rumor} used social media to study the diffusion of a rumour in the context of the 2011 Japan Earthquake, which stated that rain in the aftermath might include harmful chemical substances and led to people being warned to carry an umbrella. The authors looked at retweets of early tweets reporting the rumour, as well as later tweets reporting that it was false. While their study showed that the appearance of later correction tweets diminished the diffusion of tweets reporting the false rumour, the analysis was limited to a single rumour and does not provide sufficient insight into understanding the nature of rumours in social media. Their case study, however, does show an example of a rumour with important consequences for society, as citizens were following the latest updates with respect to the earthquake in order to stay safe.

Some researchers have looked at how social media users support or deny rumours in breaking news stories but results are, as yet, inconclusive. In some cases it has been suggested that Twitter does well in debunking inaccurate information thanks to self-correcting properties of crowdsourcing as users share opinions, conjectures, and evidence. For example, Castillo et~al. \cite{castillo2013predicting} found that the ratio between tweets supporting and debunking false rumours was 1:1 (one supporting tweet per debunking tweet) in the case of a 2010 earthquake in Chile. Procter et al. \cite{procter2013reading} came to similar conclusions in their analysis of false rumours during the 2011 riots in England, but they noted that any self-correction can be slow to take effect. In contrast, in their study of the 2013 Boston Marathon bombings, Starbird et~al. \cite{starbird2014rumors} found that Twitter users did not do so well in distinguishing between the truth and hoaxes. Examining three different rumours, they found the equivalent ratio to be 44:1, 18:1 and 5:1 in favour of tweets supporting false rumours. In a similar way, but expanding on the findings of earlier work, we aim to look at rumours associated with a broader set of newsworthy events, differentiating also the support and denial that comes before or after a rumour is resolved as being true or false.

Friggeri et~al. \cite{friggeri2014rumor} and Hannak et~al. \cite{hannak2014get} used a different approach to look at how people respond to false information by looking at tweets that link to web pages that debunk certain pieces of information. While their studies are of a broader nature and cover rumours within different contexts, they only looked at false rumours that had been debunked on the fact-checking website Snopes\footnote{http://www.snopes.com/}. Moreover, they did not look into conversational characteristics around rumours, beyond looking at users who respond with a link to Snopes.

Going beyond studying the nature of rumours to gain insight into how they are spread and supported by recipients, some researchers have also worked on the development of automated techniques to categorise the type of support provided in each tweet towards the underlying rumour \cite{lukasik2015classifying,qazvinian2011rumor}. The dataset introduced in the current work provides a suitable resource to expand research in this direction.

\subsection*{Motivation and Contribution}

%Social media have become ubiquitous as a source of information about ongoing events and breaking news. However, this rich source of information is rife with unverified and inaccurate content, which requires all users to handle information carefully to avoid  spreading inaccurate and potentially harmful information. \todo{[I deleted some of this as I think it is repetitive. Perhaps we also need to delete the above and start straight away with the following paragraph]}

In the absence of research that takes a broad look at the diffusion, support and denial of rumours in social media, we set out to develop a methodology that would enable us to collect, annotate and analyse a large collection of rumours that we could then analyse for how people react to them. In contrast to previous approaches, our methodology enables gathering collections of rumours that are not necessarily known \textit{a priori}, so that we can collect a diverse set of rumours with different levels of popularity. We introduce a scheme for the annotation of social media, which we apply to a collection of 330 conversation threads discussing different rumours, associated with 9 different newsworthy events. Using this dataset, we analyse rumour diffusion, support and denial in social media, looking at how they evolve from the point of being released to eventual resolution. Our study shows interesting differences between true and false rumours, one of them being that the former are typically resolved much faster than the latter. We find that social media users generally show a tendency towards supporting rumours whose veracity is yet to be resolved, which questions the validity of using the aggregated judgments of users for determining the veracity of a rumour. Users do reasonably well in distinguishing true and false rumours after they are corroborated or debunked, but aggregated judgments while a rumour is still unverified leave much to be desired. This emphasises the need for developing machine learning techniques that can provide assistance in assessing rumour veracity in real-time. The main contributions of our study include:

\begin{itemize}
 \item The development of an annotation scheme for analysing conversations around rumours in social media. This annotation scheme has been developed to capture observable conversational features of how people engage with information of questionable veracity.
 \item The application of this annotation scheme to a collection of 330 threads associated with 9 different newsworthy events. We tested and validated an annotation methodology both with an expert panel as well as through crowdsourcing. We came up with a crowdsourcing methodology for the creation of the annotated dataset described and analysed here.
 \item The performance of a quantitative analysis of rumours in social media, looking at how they are spread, as well as how users support or deny them. To the best of our knowledge, our work is the first to perform such analysis on a broad dataset involving multiple newsworthy events and rumourous stories, and the first to consider the change in veracity status (unverified, true, false) at different points of the life cycle of a rumour.
\end{itemize}

\section*{Materials and Methods}

The way the data is collected as well as the manual annotation process constitute an important part of our methodology, allowing us to analyse conversational features of social media.

\subsection*{Data Collection}

Previous research has focused on collecting data for rumours known to have been circulating \textit{a priori} \cite{qazvinian2011rumor,procter2013reading,starbird2014rumors} rather than a more general approach to the collection of new rumours. By gathering rumours known \textit{a priori}, one can, for instance, search for tweets with specific keywords, e.g. `London Eye fire', to retrieve tweets associated with the rumour that the London Eye had been set on fire in the context of the 2011 England riots. However, this approach will fail to identify rumours associated with events for which specific keywords have not been previously defined. In this study, our goal has been to extend rumour collection to encompass different types of rumourous stories that are not necessarily known \textit{a priori}. We did this by collecting tweets from the streaming API relating to newsworthy events that could potentially prompt the initiation and propagation of rumours. Selected rumours were then captured in the form of conversation threads. We performed the collection of rumourous conversations in three steps. First, we collected candidate rumourous stories (signaled by highly retweeted tweets associated with newsworthy current events) from Twitter. Second, journalists in the research team selected from these candidate rumours those that met the rumour criteria \cite{zubiaga2015crowdsourcing} and identified the tweets that actually introduced them. Finally, we collected associated conversations for these rumour-introducing tweets and annotated them using an annotation scheme specifically crafted for this purpose.

\subsubsection*{Identification of Rumourous Stories and Conversations}

\noindent \textbf{Data collection.} We use Twitter's streaming API to collect tweets in two different situations: (1) breaking news that are likely to spark multiple rumours and (2) specific rumours that are identified \textit{a priori}. We collected tweets for nine different events, which include five cases of breaking news: 

\begin{itemize}
 \item \textbf{Ferguson unrest:} citizens of Ferguson in Michigan, USA, protested after the fatal shooting of an 18-year-old African American, Michael Brown, by a white police officer on August 9, 2014.
 \item \textbf{Ottawa shooting:} shootings occurred on Ottawa's Parliament Hill in Canada, resulting in the death of a Canadian soldier on October 22, 2014.
 \item \textbf{Sydney siege:} a gunman held hostage ten customers and eight employees of a Lindt chocolate café located at Martin Place in Sydney, Australia, on December 15, 2014.
 \item \textbf{Charlie Hebdo shooting:} two brothers forced their way into the offices of the French satirical weekly newspaper Charlie Hebdo in Paris, killing 11 people and wounding 11 more, on January 7, 2015.
 \item \textbf{Germanwings plane crash:} a passenger plane from Barcelona to D\"usseldorf crashed in the French Alps on March 24, 2015, killing all passengers and crew. The plane was ultimately found to have been deliberately crashed by the co-pilot of the plane.
\end{itemize}

And four specific rumours (known a priori):

\begin{itemize}
 \item \textbf{Prince to play in Toronto:} a rumour started circulating on November 3, 2014 that the singer Prince would play a secret show in Toronto that night. Some people even queued at the venue to attend the concert, but the rumour was later proven false.
 \item \textbf{Gurlitt collection:} a rumour in November 2014 that the Bern Museum of Fine Arts was going to accept a collection of modernist masterpieces kept by the son of a Nazi-era art dealer. The museum did end up accepting the collection, confirming the rumours.
 \item \textbf{Putin missing:} numerous rumours emerged in March 2015 when the Russian president Vladimir Putin did not appear in public for 10 days. He spoke on the 11th day, denying all rumours that he had been ill or was dead.
 \item \textbf{Michael Essien contracted Ebola:} a post by a Twitter user on October 12, 2014 stated that the AC Milan footballer Michael Essien had contracted Ebola. The report was later denied by the footballer and thus exposed as a hoax.
\end{itemize}

Collection through the streaming API was launched straight after the journalists identified a newsworthy event likely to give rise to rumours. As soon as the journalists informed us about a newsworthy event, we set up the data collection process for that event, tracking the main hashtags and keywords pertaining to the event as a whole. Twitter's public stream was accessed through the 'statuses/filter' endpoint\footnote{\url{https://dev.twitter.com/streaming/reference/post/statuses/filter}}, making use of the 'track' parameter to specify keywords. Note that while launching the collection slightly after the start of the event means that we may have missed the very early tweets, we kept collecting subsequent retweets of those early tweets, making it much more likely that we would retrieve the most retweeted tweets from the very first minutes. This is also consistent with our methodology as we rely on highly retweeted tweets to identify candidate rumours.

\noindent \textbf{Data sampling.} Given the large volume of the datasets, we sampled rumour candidates by picking tweets that sparked a high number of retweets. This is justified by our definition of rumours, which specifies that a statement has to generate substantial interest and be widely circulated in order to be considered as a rumour. Depending on the event, the retweet threshold used ranged from two for specific rumours (Prince, Gurlitt, Putin, Michael Essien) to up to 100 for larger-scale events (Ferguson, Ottawa, Sydney, Charlie Hebdo). Although the crash of the Germanwings airplane in the French Alps was a significant event that generated Twitter conversations over the course of several days, a lower threshold of 25 retweets was chosen in order to capture an adequate number of tweets in German. The sampled subsets of tweets were visualised in a timeline with links to the conversation thread sparked by each of the initial (source) tweets. Next, journalists were asked to identify rumourous source tweets and non-rumourous source tweets. Journalists often need to clarify information they need to use and thus have experience separating corroborated facts from rumours. Additionally, the journalists continually tracked the nine events, so they were knowledgeable about the stories and rumours associated with them.

\noindent \textbf{Data annotation.} We developed an annotation tool to facilitate the journalists' work of annotating tweets as rumours vs non-rumours. The visualisation of source tweets in the form of a timeline, accompanied by the context added by the conversation thread they spark, made it easier to keep track of posts as the event unfolds. While data collection was concurrent with events unfolding, the annotation process was performed \textit{a posteriori}, once a consensus had emerged about the facts relating to the event in question. This enabled the journalists to be more knowledgeable about the entire story, as well as any rumours that had emerged and how they had been resolved. It also encouraged careful annotation that encompassed a broad set of rumours; the journalists could go through the whole timeline of tweets as we presented them and perform the annotations. The annotation work performed by the journalists was twofold: (a) annotation of each source tweet as being a rumour or not and (b) grouping of source tweets into stories; the latter helped us group different rumourous tweets (and associated conversation threads) as being part of the same story, as is the case with the following three tweets which are all associated with the same rumour (i.e., associated with the story ``Putin is facing a palace coup''):

\begin{itemize}
 \item Is Putin facing palace coup? [link]
 \item Can Putin's Absence Indicate A Palace Coup In Moscow? \#europe \#russia [link]
 \item RUMORS about \#Putin started with a stroke, then he's sick, and now this [link] What's next... \#Russia \#cdnpoli \#dirtydeeds
\end{itemize}

\begin{figure}[tbh]
  \begin{center}
   \includegraphics[width=1.0\textwidth]{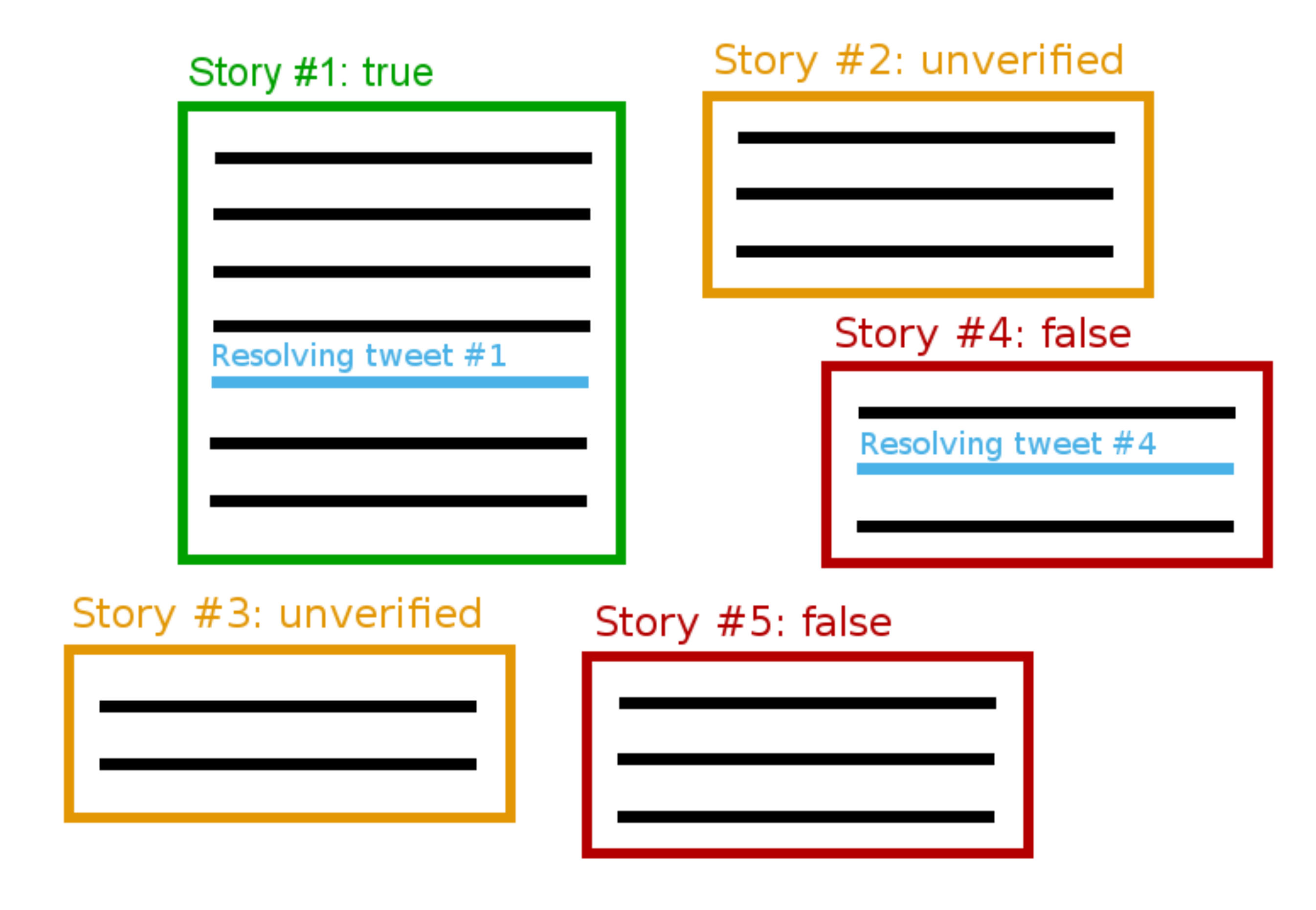}
   \caption{Diagram showing the structure resulting from the journalists' annotation work. Rumour stories, represented by squares, can be one of true (green), false (red), or unverified (orange). Each of the rumour stories has a number of rumour threads associated with it (black lines). When a story is true or false, the journalists also picked, where available, one tweet as the resolving tweet within the story's timeline.}
   \label{fig:annotation-structure}
  \end{center}
\end{figure}

Once rumours were grouped into stories, the journalists also annotated whether each of the stories was ultimately proven to be false, confirmed as true, or remained unverified. For cases where the story was either true or false, they also marked within the timeline of the story the tweet that, according to their judgement, was decisive for establishing the veracity of the rumour. We refer to the latter as the \textit{resolving tweet}. Note that the resolving tweet is not always available for true and false stories, as such information may sometimes occur outside of our data collection time frame, e.g. weeks later or outside the Twittersphere. Fig \ref{fig:annotation-structure} shows a diagram with the structure of rumour data produced with the annotation work performed by the journalists.

Our datasets consist of rumour stories, represented by squares, which can be one of true (green), false (red), or unverified (orange). Each of the rumour stories has a number of rumour threads associated with it, which we represent as black lines that form a timeline where threads are sorted by time. When a story is true or false, the journalists also picked, within the story's timeline, one tweet as the resolving tweet. Note that the resolving tweets cannot always be found within the Twitter timeline, as in example story \#5.

\noindent \textbf{Outcome of the rumour annotation task.} Table \ref{tab:rumour-annotation-stats} summarises the outcome of the annotation work, showing the number of source tweets annotated as rumours and non-rumours for each of the nine events, as well as the number of rumourous stories, where each story corresponds to a group of rumourous tweets.

\begin{table}
 \begin{adjustwidth}{-0.5in}{0in}
  \footnotesize
  \centering
  \begin{tabular}{| l | c | c | c | c |}
   \hline
   Event name & Rumour stories & Annotated threads & Rumour threads & Non-rumour threads \\
   \hline
   Sydney Siege & 61 & 1321 & 535 & 786 \\
   \hline
   Ottawa Shooting & 51 & 901 & 475 & 426 \\
   \hline
   Charlie Hebdo & 61 & 2169 & 474 & 1695 \\
   \hline
   Germanwings & 19 & 1022 & 332 & 690 \\
   \hline
   Ferguson & 42 & 1183 & 291 & 892 \\
   \hline
   Prince to play in Toronto & 6 & 241 & 237 & 4 \\
   \hline
   Gurlitt & 3 & 386 & 190 & 196 \\
   \hline
   Putin missing & 6 & 266 & 143 & 123 \\
   \hline
   Essien has Ebola & 1 & 18 & 18 & 0 \\
   \hline					
   TOTAL & 250 & 7507 & 2695 & 4812 \\
   \hline
  \end{tabular}
  \caption{Outcome of the annotation of rumours.}
  \label{tab:rumour-annotation-stats}
 \end{adjustwidth}
\end{table}

For the purposes of our study, we focus on the 2,695 tweets annotated as rumourous. These rumourous tweets have been annotated in three different languages: 2,460 in English, 198 in German, and 37 in French.

\noindent \textbf{Complementing the Dataset with Conversations.} We used the above dataset of rumourous source tweets to launch collection of the conversational threads they initiated. As a native feature of Twitter, users can reply to one another. Hence, we look for all the replies to the 2,695 rumourous source tweets for the nine events in the dataset. While Twitter does not provide an API endpoint to retrieve conversations sparked by tweets, it is possible to collect them by scraping tweets through the web client interface. We developed a script that enabled us to collect and store complete conversations for all the rumourous source tweets\footnote{The conversation collection script is available at \url{https://github.com/azubiaga/pheme-twitter-conversation-collection}}. The script scrapes the Twitter web client interface, collecting the responses that appear beneath the source tweet. Once the tweets below the source tweet have been scraped, the script performs the following two steps to make sure that the whole conversation started by the source tweet is retrieved: (1) The script checks if there are more pages with responses (since Twitter pages the responses), and (2) the script then retrieves, recursively, the replies to all those replying tweets, which enables retrieval of nested interactions. This process gives us the tweet IDs of all the tweets replying to the source tweet, i.e. the conversational thread, from which we can form the whole tree. To collect all the metadata for those tweets, we then access the Twitter API using the tweet IDs; specifically, we use the 'statuses/lookup' endpoint\footnote{\url{https://dev.twitter.com/rest/reference/get/statuses/lookup}}, which provides all the metadata for up to 100 tweets at once.

\begin{figure}[tbh]
  \begin{center}
   \includegraphics[width=1.2\textwidth]{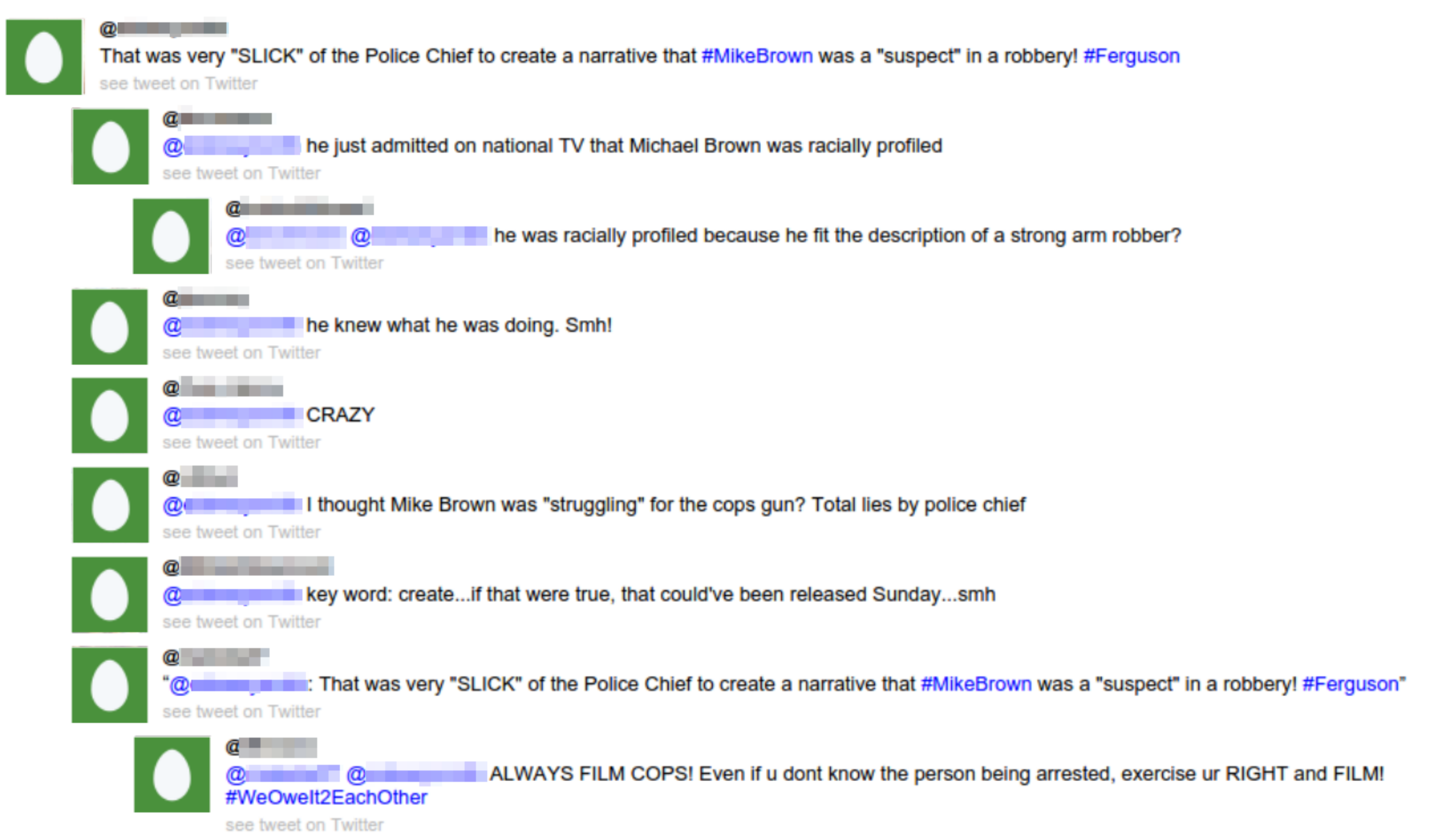}
   \caption{Example of a conversation generated by a rumourous tweet.}
   \label{fig:rumourous-conversation}
  \end{center}
\end{figure}

The collection of conversations for the 2,695 rumourous source tweets contained 34,849 tweets, 2,695 being source tweets and 32,154 being responses.

\subsubsection*{Annotation of Rumourous Conversations}

Our ultimate goal is to have rumourous conversations annotated according to the support and evidence provided by tweets responding to the rumour denoted by the source tweet (response tweets). We developed an annotation scheme suitable for capturing conversation properties of the Twitter threads in terms of such interactions and used it to obtain an annotated corpus using crowdsourcing.

In designing the annotation scheme we considered existing schemes for factuality and aspects of conversation analysis, such as turn-taking and corresponding reactions. The development of the annotation scheme has been motivated by previous research that has evolved conversation analytic techniques in order to arrive at an appropriate methodology for the sociological analysis of microblogs, especially Twitter \cite{tolmie2015arXiv151103193T}. This research identified a number of core organisational properties within Twitter conversations that have formed the basis of the approach reported here for identifying potentially rumourous conversation threads, i.e.: that it takes more than one conversational turn for a rumour to be identified; that Twitter conversations have a sequential order; that they involve topic management; and that the production of a response is shaped by its accountable characteristics. 
The development of the annotation scheme was carried out in two iterations: first by testing it with a four-person expert panel of social media researchers (three PhD students, one postdoctoral researcher) and second by running preliminary tests on a crowdsourcing platform. The annotation scheme resulting from this iterative process is shown in Fig \ref{fig:annotation-scheme}. It assumes that there are two types of tweets in a conversation: (1) the source tweet that initiates a rumour, and (2) response tweets that respond within the thread started by the source tweet, including both direct and nested responses. The annotation scheme consists of three main dimensions which express the mode of interaction:
(1) support or response type, (2) certainty and (3) evidentiality. The annotation of source tweets and response tweets presents two variants in the case of the first component; support applies to source tweets whereas response types applies to response tweets. Each tweet in a rumourous conversation is annotated in terms of the above three dimensions, which we define in detail below:

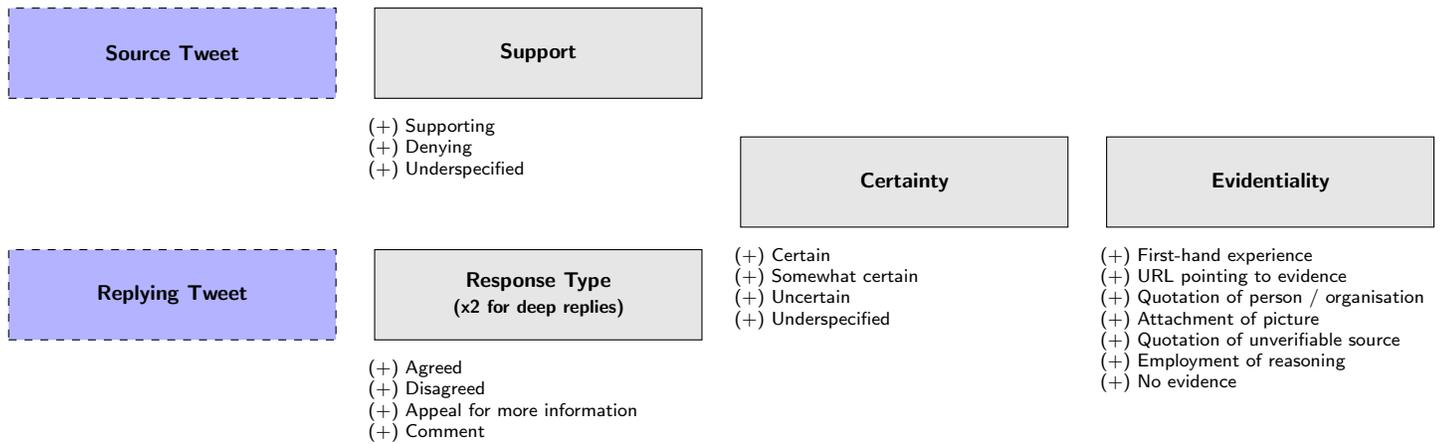
\begin{figure*}[tbh]
\begin{adjustwidth}{-1.25in}{0in}
\centering
\begin{tikzpicture}
[node distance = 1cm, auto,font=\footnotesize,
every node/.style={node distance=3cm},
comment/.style={rectangle, inner sep=0pt, text width=4.5cm, node distance=0.25cm, font=\scriptsize\sffamily},
box/.style={rectangle, draw, fill=black!10, inner sep=5pt, text width=4cm, text badly centered, minimum height=1.2cm, font=\bfseries\footnotesize\sffamily},
titlebox/.style={rectangle, draw, fill=blue!30, inner sep=5pt, text width=4cm, text badly centered, minimum height=1.2cm, font=\bfseries\footnotesize\sffamily}]

% Source tweets, boxes
\node [box] (support) {Support};
\node [box, below right=0.5cm and 0.5cm of support] (certainty1) {Certainty};
\node [box, right=0.5cm of certainty1] (evidentiality1) {Evidentiality};
\node [titlebox, dashed, left=0.5cm of support] (source) {Source Tweet};

% Replying tweets, boxes
\node [box, below=2cm of support] (responsetype) {Response Type \\ \scriptsize(x2 for deep replies)};
\node [titlebox, dashed, left=.5cm of responsetype] (reply) {Replying Tweet};

% SUPPORT
\node [comment, below=0.25cm of support] {(+) Supporting\\
(+) Denying\\
(+) Underspecified};

% EVIDENTIALITY (Source)
\node [comment, below=0.25 of evidentiality1] {(+) First-hand experience\\
(+) URL pointing to evidence\\
(+) Quotation of person / organisation\\
(+) Attachment of picture\\
(+) Quotation of unverifiable source\\
(+) Employment of reasoning\\
(+) No evidence};

% CERTAINTY (Source)
\node [comment, below=0.25 of certainty1] (comment-certainty1) {(+) Certain\\
(+) Somewhat certain\\
(+) Uncertain\\
(+) Underspecified};

% RESPONSE TYPE
\node [comment, below=0.25cm of responsetype] {(+) Agreed\\
(+) Disagreed\\
(+) Appeal for more information\\
(+) Comment};

 \end{tikzpicture}

 \caption{Annotation scheme for rumourous social media conversations.}
 \label{fig:annotation-scheme}
\end{adjustwidth}
\end{figure*}

\begin{enumerate}
 \item \textbf{Support and Response Type:}

 \begin{itemize}
  \item \textbf{Support:} Support is only annotated for source tweets. It defines if the message in a source tweet is conveyed as a statement that supports or denies the content of the statement. It is hence different from the rumour's truth value and reflects the view of the author of the source tweets towards the rumour's veracity. Support takes the following values: (1) \emph{Supporting}, when the author of the source tweet supports the content of the statement, (2) \emph{Denying}, when denying it or (3) \emph{Underspecified}, when the author's view is unclear. This feature is related to the ``Polarity'' feature in the factuality scheme by Saur\'i et al. \cite{sauri2009factbank}.

  \item \textbf{Response Type:} Response Type is used to designate the support of response tweets towards a source tweet that introduces a rumourous story. Some replies can be very helpful in determining the veracity of a rumour, and thus we annotate Response Type with one of the following four values: (1) \emph{Agreed}, when the author of the reply supports the statement they are replying to, (2) \emph{Disagreed}, when the author of the reply disagrees with the statement they are replying to, (3) \emph{Appeal for more information}, when the author of the reply asks for additional evidence in relation to the statement they are responding to, or (4) \emph{Comment}, when the author of the reply makes their own comment without a clear contribution to assessing the veracity of either the tweet they are responding to or the source tweet. The inclusion of the Response Type dimension in the annotation scheme follows Procter et al. \cite{procter2013reading}, who originally introduced these four types of responses for rumours. However, unlike \cite{procter2013reading} we additionally consider the annotation of Response Type for nested replies, i.e. tweets that are not directly replying to the source tweet. In this case Response Type is annotated for two different aspects: (i) the type of response with respect to the rumour in the source tweet and (ii) the type of response towards its parent tweet, i.e. the tweet it is directly replying to. This double annotation allows us to better analyse the way conversations flow and how opinions evolve with respect to veracity. It is worth noting that the response type is not necessarily transitive, and the aggregation of pairwise agreements and disagreements with previous tweets does not necessarily match with the agreement with the source.
  
  For the purposes of our analysis here, we simplify a tweet's response type as its support with respect to the rumour; we infer this from the tweet's support with respect to the source, which can be one of these four cases: (1) if the source tweet supports a rumour and the response tweet agrees, we consider that the tweet supports the rumour, (2) if the source tweet supports a rumour and the response tweet disagrees, we consider that the tweet denies the rumour, (3) if the source tweet denies a rumour and the response tweet agrees, we consider that the tweet denies the rumour, and (4) if the source tweet denies a rumour and the response tweet disagrees, we consider that the tweet supports the rumour. Figure \ref{fig:support-annotation-example} shows an example of the double annotation, and how we determine the support towards the rumour.

  \begin{figure}[ht]
    \centering
    \includegraphics[width=1.0\textwidth]{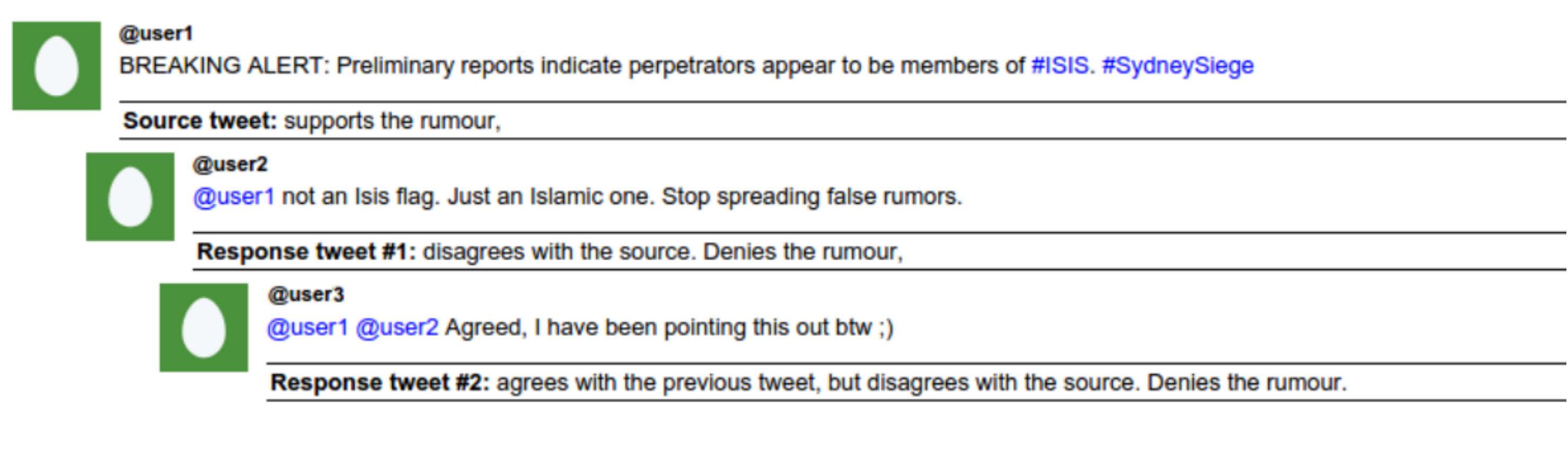}
    \caption{Example of annotation of rumour type, as well as how we determine support towards the rumour.}
    \label{fig:support-annotation-example}
  \end{figure}
 \end{itemize}

 \item \textbf{Certainty:} Certainty measures the degree of confidence expressed by the author of a tweet when posting a statement in the context of a rumour. It applies to both source tweets and response tweets. The author can express different degrees of certainty when posting a tweet, from being 100\% certain, to considering it as a dubious or unlikely occurrence. The value annotated for either Support or Response Type has no effect on the annotation of Certainty, and thus it is coded regardless of the statement supporting or denying the rumour. The values for Certainty include: (1) \emph{Certain}, when the author is fully confident or the author is not showing any kind of doubt, (2) \emph{Somewhat certain}, when they are not fully confident and (3) \emph{Uncertain}, when the author is clearly unsure. This corresponds to the Modality feature of  Saur\'i et al. \cite{sauri2009factbank}.

 \item \textbf{Evidentiality:} Evidentiality determines the type of evidence (if any) provided by an author of a tweet and applies to both source tweets and response tweets. It is important to note that the evidence provided has to be directly related to the rumour being discussed in the conversation and any other kind of evidence that is irrelevant in that context is not annotated here. Evidentiality can have the following values: (1) \emph{First-hand experience}, when the author of the tweet claims to have witnessed events associated with the rumour (2) \emph{Attachment of a URL} pointing to evidence, (3) \emph{Quotation} of a person or organisation, when an accessible source is being quoted as a source of evidence, (4) \emph{Attachment of a picture}, (5) Quotation of an \emph{unverifiable source}, when the source being mentioned is not accessible, such as ``my friend said that...'', (6) \emph{Employment of reasoning}, when the author explains the reasoning behind their view and (7) \emph{No evidence}, when none of the other types of evidence is given in the tweet. Contrary to the dimensions described above, more than one value can be picked for Evidentiality, with the exception of ``No evidence'' which excludes the selection of any other value. Hence, we cater for the fact that a tweet can provide more than one type of evidence, e.g. quoting a news organisation while also attaching a picture that provides evidence.
\end{enumerate}

All three of the above dimensions are annotated for each tweet in a conversation thread, with one exception; response tweets whose Response Type has been annotated as Comment, do not need to be annotated for Certainty and Evidentiality.

Since the annotation of all 2,695 conversations we collected was not viable due to time and budget constraints, we randomly sampled a subset of the data. The sampled dataset includes 330 rumourous conversations, 297 of which are in English and 33 in German. This amounts to 4,842 tweets overall, 4,560 in English and 282 in German. The conversations are distributed across the nine events as follows: Charlie Hebdo (74), Sydney siege (71), Ottawa shooting (58), Germanwings crash (53), Ferguson unrest (46), Putin missing (13), Prince/Toronto (12), Ebola/Essien (2), and Gurlitt (1).

These 330 conversations were categorised into 140 different stories. Out of these 330 conversations, 159 are true, 68 are false and 103 remained unverified. 132 of the conversations were part of one of the 41 stories that have a resolving tweet annotated.

The annotation of the sampled rumourous conversation threads we collected was performed through crowdsourcing in order to ensure its timely completion \cite{Procter2013enabling}. For crowdsourcing, we used the CrowdFlower platform\footnote{http://www.crowdflower.com/}, because of its flexibility, which allowed us to refine the annotation work and specify prerequisites for annotators. Further details on how we split the annotation into individual tasks and the settings chosen for the crowdsourcing tasks can be found in \cite{zubiaga2015crowdsourcing}, where we validated the suitability of the approach, including both the annotation scheme, as well as the crowdsourcing jobs, on a small subset of rumours.

The annotation of all 330 conversation threads consisted of 68,247 judgments on 4,842 tweets performed by 233 different annotators. We combined all judgments for each tweet-dimension pair through majority vote. These tweets are annotated for the three features described in our annotation scheme: support or response type (which is annotated twice for deep replies), certainty, and evidentiality. A comparative analysis between the three features shows that they are independent from each other, with low pairwise Pearson correlation values: -0.31 between support and certainty, -0.15 between support and evidentiality, and -0.04 between certainty and evidentiality. To quantify the difficulty of the task, we measured the inter-annotator agreement values by comparing each individual judgment against the majority vote. Overall, the annotators achieved an agreement rate of 62.3\%, which is distributed differently across different tweet types and dimensions. Table \ref{tab:agreement-rates-by-feature} shows how the agreement rates are distributed for source and response tweets when annotating for Support, Certainty and Evidentiality. Our findings show that annotators found it easier to annotate source tweets, as they are less ambiguous and require less context for understanding. The agreements are somewhat lower for response tweets. When we compare the different dimensions, we observe that Support is the easiest to annotate for source tweets, but very similar to Certainty overall. Evidentiality is slightly more difficult to annotate for both source and response tweets, presumably because of the large number of different values that the annotators can choose from.

\begin{table}[htb]
 \centering
 \begin{tabular}{| l || c | c | c |}
  \hline
  & Support & Certainty & Evidence \\
  \hline
  \hline
  Source tweets & 81.1\% & 68.8\% & 74.9\% \\
  \hline
  Replies & 62.2\% & 59.8\% & 58.3\% \\
  \hline
  \hline
  Overall & 63.7\% & 61.1\% & 60.8\% \\
  \hline
 \end{tabular}
 \caption{Inter-annotator agreement values for different features and tweet types.}
 \label{tab:agreement-rates-by-feature}
\end{table}

The present study was approved by the Warwick University Humanities \& Social Sciences Research Ethics Committee (HSSREC), which confirmed that the project, ``Computing Veracity Across Media, Languages and Social Networks (PHEME)'', received full ethics approval (ref 69/13-14, date: 30.05.2014), including approval to publish extracts from social media datasets.

The resulting dataset is available for research purposes\footnote{\url{https://figshare.com/articles/PHEME_rumour_scheme_dataset_journalism_use_case/2068650}}. We believe that the methodology described above, along with the conversation collection software, should enable reproducibility of similar data collections for future events.
% \todo{The resulting dataset is available for research purposes at...}
% Link above to be added with the final paper, we can share a private link with reviewers

\section*{Results}

We begin by investigating the diffusion of rumours in our corpus and then move on to analyse the annotations for Support, Certainty, and Evidentiality for both source and response tweets. We conclude by exploring attributes of users that participate in rumour diffusion.

\subsection*{Rumour Timelines}

\begin{figure}[ht]
  \centering
  \includegraphics[width=0.9\textwidth]{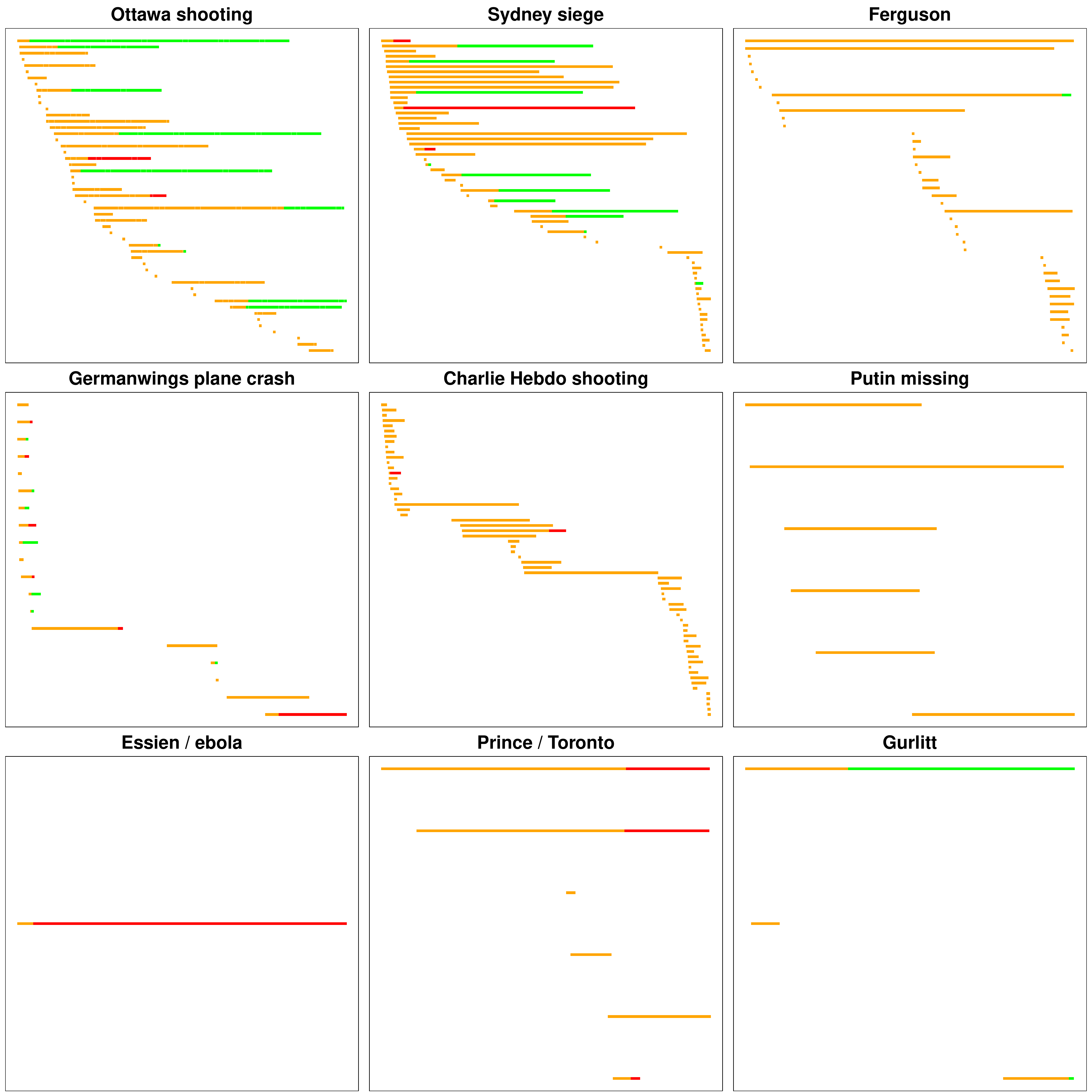}
  \caption{Rumour timelines showing the lifecycle of rumours that start as unverified stories (orange), and are occasionally later resolved as being either true (green), or false (red). Each line represents a rumour story, while the X axis represents the timeline.}
  \label{fig:rumour-timelines}
\end{figure}

Figure \ref{fig:rumour-timelines} shows the timelines for the rumours collected and annotated for the nine events in our dataset. Rumours are coloured according to their veracity status at each moment, as determined by the journalists through the annotation of resolving tweets. All of the rumours start off in an unverified status (orange), while some of them are later proven true (green) or found to be false (red); others, however, remain unverified, and are featured in orange through the entire timeline. These charts visualise, in all, 103 rumours which remain unverified, 159 which were later proven, and 68 which were found to be false. Out of the 227 rumours that turned out to be either true or false, we have a resolving tweet for 129 rumours (56.8\%), while for the other 98 cases, the journalists could not find a resolving tweet. Some of these may have been resolved outside of Twitter, outside of the collected data, or later in time out of our data collection and annotation time frames; these are coloured in orange. The figures show variability in the duration of the unverified status of rumours; some rumours are resolved quite quickly while others take much longer to be verified. They also highlight the large number of unverified rumours in the case of both the Ferguson unrest and the Charlie Hebdo shooting. On the other hand, more rumourous stories ended up being verified in events like the Ottawa shooting or the Sydney siege. While the distribution of veracity status varies across different events, the number of rumours is generally large but the number of rumours of unverified status is remarkable.

\begin{figure}[ht]
 \centering
 \includegraphics[width=0.5\textwidth]{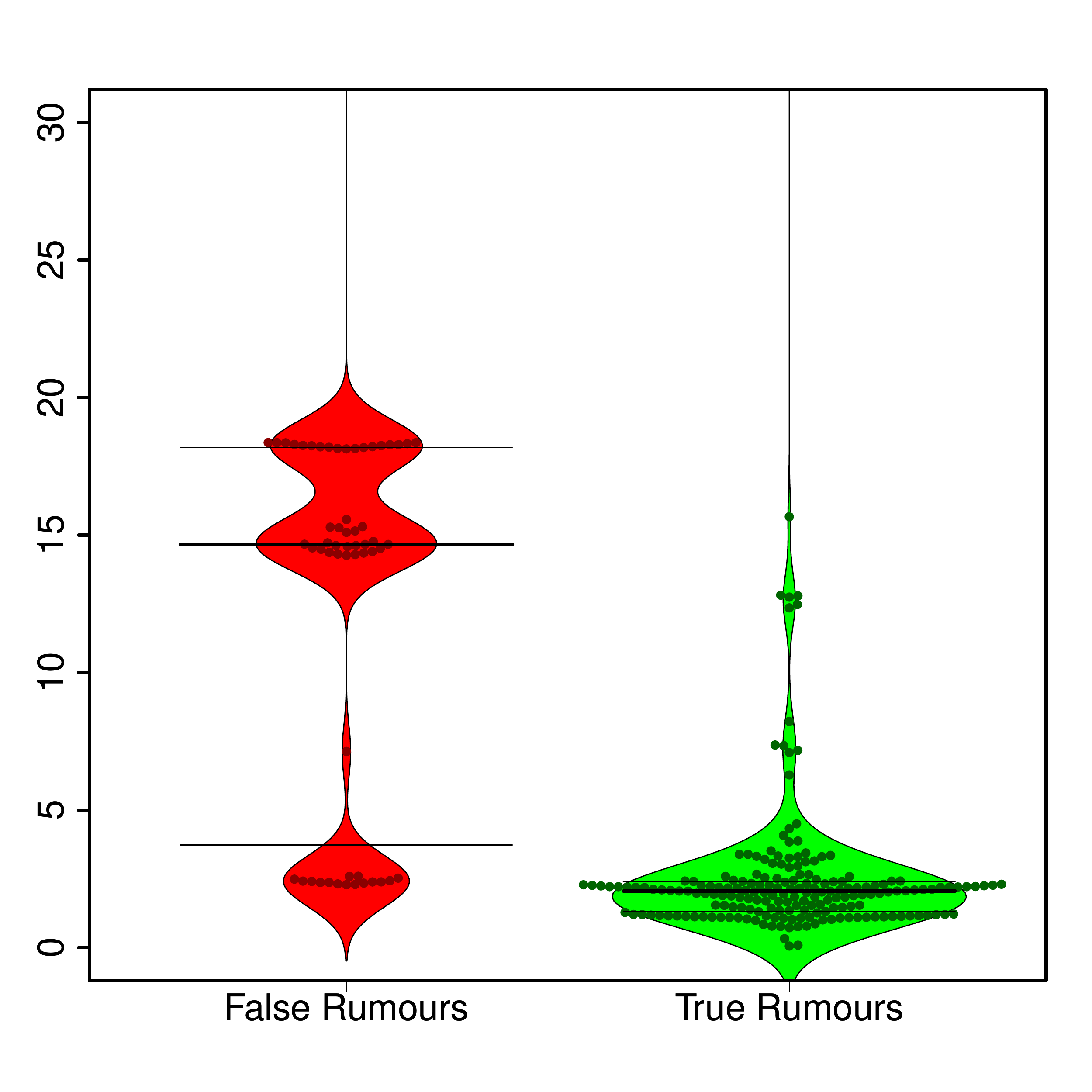}
 \caption{Distribution of delays (in hours) in resolving false (red) and true (green) rumours. Horizontal lines represent 25, 50, and 75 percentiles.}
 \label{fig:rumour-resolving}
\end{figure}

When we look at the delay between a rumour being posted for the first time on Twitter and the rumour being resolved as being either true or false, we observe that false rumours tend to be resolved significantly later than true rumours (Wilcoxon signed-rank test \cite{wilcoxon1945individual}: $W = 812.5$, $p < 2.2e^{-16}$). While the median true rumour is resolved in about 2 hours, the median false rumour takes over 14 hours to be resolved. Figure \ref{fig:rumour-resolving} illustrates this large difference in the time it takes for true and false rumours to be resolved. While the vast majority of true rumours are resolved within 5 hours of being posted for the first time, many of the false rumours are not resolved within the first 10 hours; indeed for those that are eventually solved, this does not happen until 15 to 20 hours after the timestamp of the source tweet introducing the rumour.

\subsection*{Rumour Diffusion}

\begin{figure}[ht]
  \centering
  \includegraphics[width=0.9\textwidth]{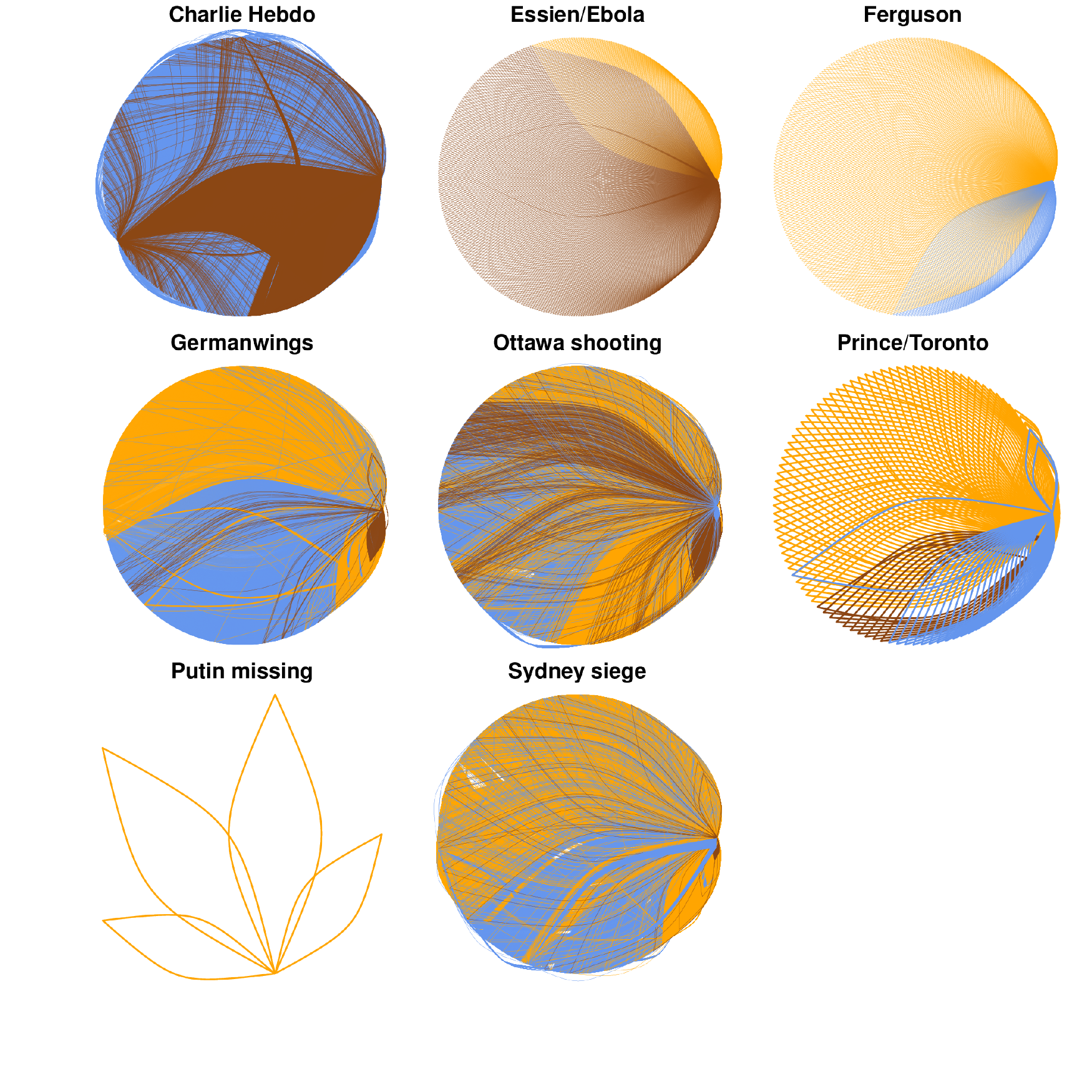}
  \caption{Retweet networks representing retweets of unverified source tweets (orange), accurate tweets that support true rumours or deny false rumours (blue) and inaccurate tweets that deny true rumours or support false rumours (brown).}
  \label{fig:retweet-networks}
\end{figure}

To complement the analysis enabled by the visualisation of rumour timelines, we take a closer look at the diffusion of these rumours in the form of retweets. Figure \ref{fig:retweet-networks} shows networks of interactions between users, the connections being coloured according to their accuracy, as we describe next. Note that eight events are shown here, given that the other event, Gurlitt, did not spark sufficient retweets to create an equivalent visualisation. Here we consider a connection between two users \textit{a,b} if a user \textit{a} retweets a user \textit{b}'s rumourous tweet. Note that retweets of rumourous source tweets are used here for the analysis, to explore how different rumours are spread. We colour these connections based on the accuracy of the original tweets being retweeted:

\begin{itemize}
 \item \textbf{Blue (accurate retweets):} the retweets of tweets that are either supporting true rumours, or denying false rumours.
 \item \textbf{Brown (inaccurate retweets):} the retweets of tweets that are wrong, i.e., either supporting false rumours, or denying true rumours.
 \item \textbf{Orange (unverified retweets):} the retweets of tweets that still have an unverified status.
\end{itemize}
 
Hence, these retweet networks visualise the degree of diffusion generated by accurate, inaccurate, and unresolved tweets. Note that each of these networks visualises all the interactions with rumours in each of the events, without considering different points in time.

In these figures, we can observe that tweets reporting unverified rumours are more widely spread; the percentages of unverified tweets range from 30.73\% of the retweets in Ebola/Essien, to 100\% of the retweets in the Putin missing story. Retweets of inaccurate reports are especially remarkable in the Ferguson unrest (26.78\%) and the Ebola/Essien hoax (69.27\%). Retweets of accurate tweets can only be observed in Prince/Toronto (23.36\%), Germanwings plane crash (33.32\%), Ottawa shootings (35.76\%), and the Sydney siege (26.64\%). Table \ref{tab:retweet-percentages} shows the percentages of retweets of each type of tweet (unverified, accurate, inaccurate) for each of the events.

\begin{table}[htb]
 \centering
 \begin{tabular}{| l || c | c | c |}
  \hline
  & Unverified & Accurate & Inaccurate \\
  \hline
  \hline
  Ebola/Essien & 30.73\% & 0.00\% & 69.27\% \\
  \hline
  Putin missing & 100.00\% & 0.00\% & 0.00\% \\
  \hline
  Charlie Hebdo & 96.22\% & 0.00\% & 3.78\% \\
  \hline
  Prince/Toronto & 67.29\% & 23.36\% & 9.35\% \\
  \hline
  Ferguson & 73.22\% & 0.00\% & 26.78\% \\
  \hline
  Germanwings & 60.24\% & 33.32\% & 6.44\% \\
  \hline
  Ottawa shootings & 55.02\% & 35.76\% & 9.23\% \\
  \hline
  Sydney siege & 71.72\% & 26.64\% & 1.64\% \\
  \hline
 \end{tabular}
 \caption{Percentages of retweets for unverified, accurate, and inaccurate tweets.}
 \label{tab:retweet-percentages}
\end{table}

\begin{figure}[ht]
 \begin{adjustwidth}{-0.8in}{0in}
  \centering
  \includegraphics[width=1.3\textwidth]{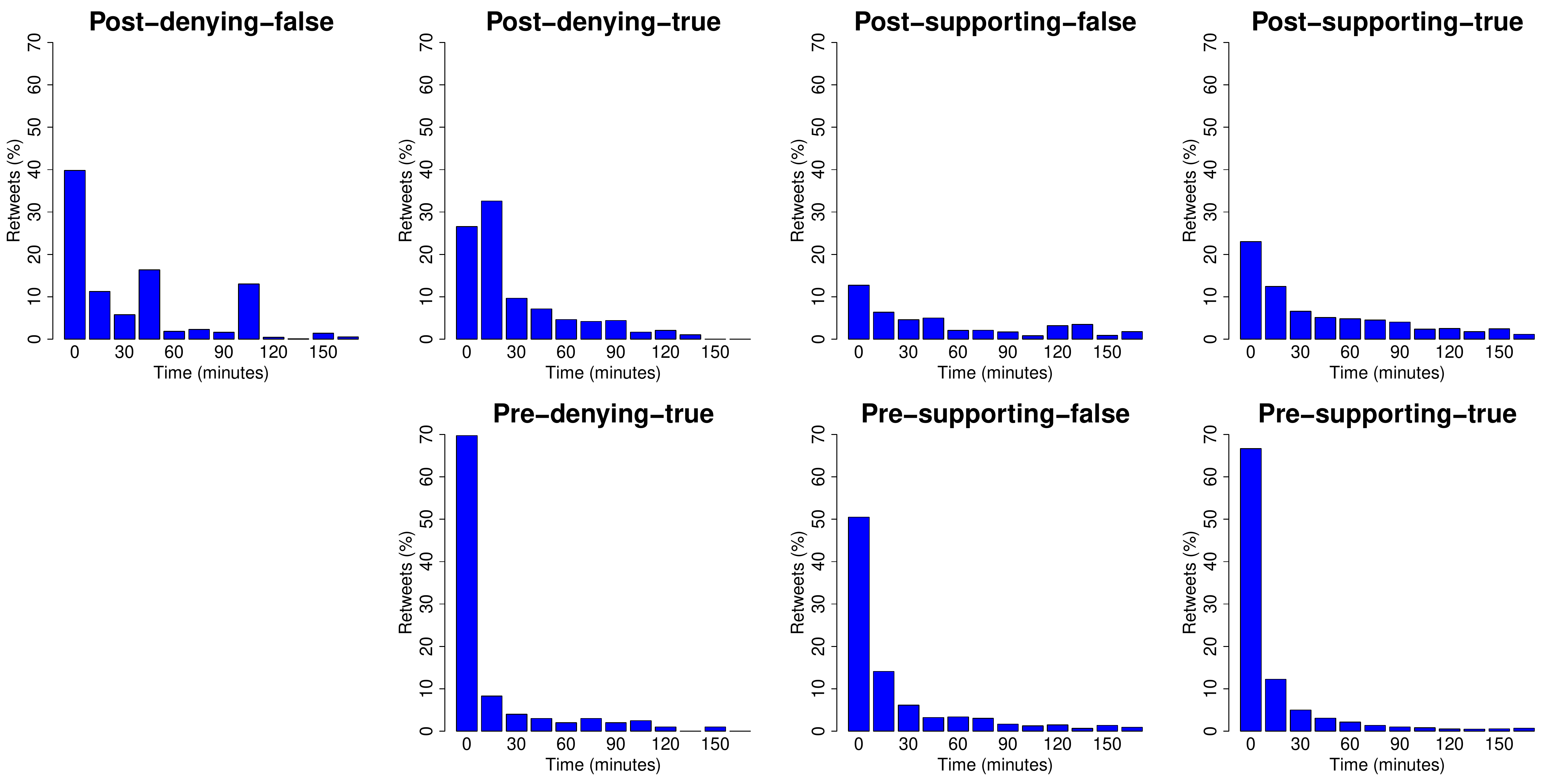}
  \caption{Retweet timelines showing the percentage of retweets that each type of tweet gets in 15 minute steps. Higher retweet percentages at the beginning represent a high interest in spreading the tweet in the very first minutes.}
  \label{fig:retweet-timelines}
 \end{adjustwidth}
\end{figure}

We further examine these retweet networks by looking at the time in which the retweets occur. We want to know the extent to which each type of rumour is retweeted, as well as whether there is a time-effect on rumour diffusion patterns, e.g. with some rumours being retweeted more at the beginning and tailing off as time passes. Figure \ref{fig:retweet-timelines} shows the average distribution of retweets for different rumourous stories (true, false) over time. The plots are normalised to show the percentages of retweets, computed every 15 minutes, representing a ratio of the total amount of retweets for that tweet. That is, once a rumourous tweet is posted, we look at the ratio of retweets that occur over time within 15 minute windows. A high percentage of retweets at a certain point in time represents that, on average, that type of source tweet receives a high number of retweets at that stage of its lifecycle. To enable a better comparison we make a distinction between tweets that occur before and after resolving tweets (post- and pre-), tweets that are part of a true or false rumour (true, false), and tweets that support or deny a rumour (supporting, denying). Note that one chart is missing, the one corresponding to the combination pre-denying-false, as there are no such instances in our annotated data.

We can observe that tweets posted before resolving tweets tend to spark a large number of retweets in the very first minutes, a trend which quickly fades in less than 20 minutes. After the occurrence of resolving tweets, retweets are more evenly distributed over time, which reveals that post-tweets keep being retweeted for a longer time. Hence, early tweets before a rumour is settled are highly spread as soon as they are posted; however, even after a rumour is resolved the rumour-bearing source tweets are retweeted for a while after.

\begin{figure}[ht]
  \centering
  \includegraphics[width=0.5\textwidth]{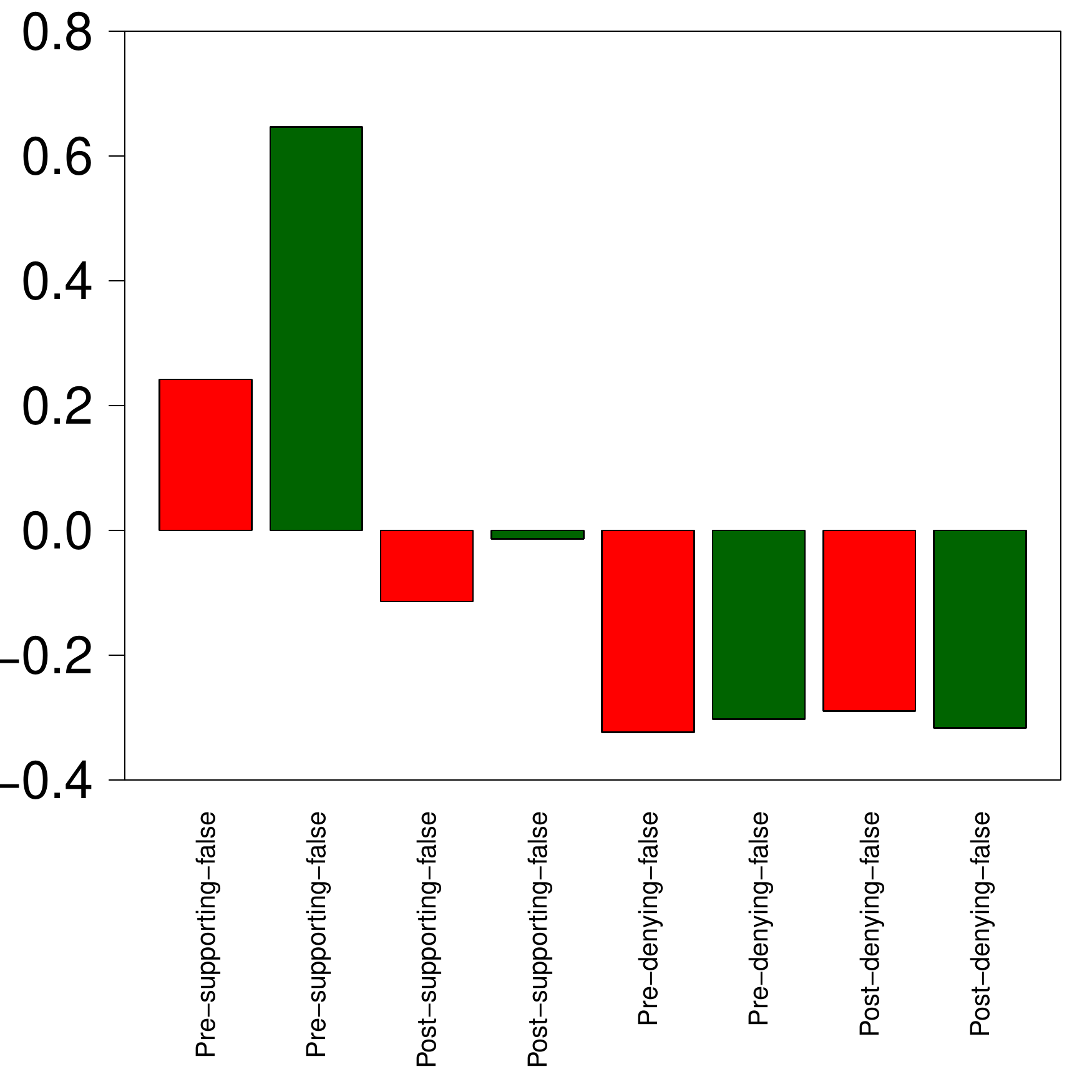}
  \caption{Z-score values for retweets of different types of tweets, representing the extent to which different types of tweets receive retweets below or above the overall average of retweets that all rumours get.}
  \label{fig:retweet-zscores}
\end{figure}

Extending this analysis, Figure \ref{fig:retweet-zscores} compares the number of retweets that each type of tweet receives overall, compared with the average number of retweets that all rumours get. The point of this check is to establish if certain types of tweets spark more retweets. Given that different events and rumours garner very different numbers of retweets, here we use z-scores to normalise the retweet counts across events and rumours. The $z$ score for a given value $x$ is given by the formula:

\begin{equation}
 z = \frac{x - \mu}{\sigma}
\end{equation}

where $\mu$ is the mean of the population, and $\sigma$ is the standard deviation of the entire population under consideration. The z-score represents, therefore, the number of standard deviations that a population is above or below the global average.

Interestingly, the types of tweets that spark more retweets are, by far, early tweets supporting a rumour that is still unverified (``pre-supporting'') ($W = 118760$, $p < 2.2e-16$). Among these, there is also a significant difference between true and false rumours; those that end up being true are retweeted more than the false ones ($W = 6145.5$, $p = 0.02678$). The number of retweets of tweets supporting these stories (``post-supporting''), both true and false, drops dramatically after the rumours are resolved, suggesting that the rumour's interest decreases once its veracity value is known. The tweets that are retweeted the least, however, are those denying a rumour, irrespective of whether the stories are true or false and whether or not they have been corroborated. This analysis reveals an interesting pattern in behaviour, showing that users tend to support unverified rumours (whether explicitly or implicitly), potentially due to the arousal that these early, unverified stories produce and their potential societal impact. Users do not seem to make the same effort, however, to spread false rumour debunks to let others know about false reports.

% \todo{Pre-supporting-retweets (false): 0.24218047654007}
% \todo{Post-supporting-retweets (false): -0.11398982064005}
% \todo{Pre-denying-retweets (false): -0.32368390045776}
% \todo{Post-denying-retweets (false): -0.28979568575383}
% \todo{Pre-supporting-retweets (true): 0.6467448558214}
% \todo{Post-supporting-retweets (true): -0.013779192808454}
% \todo{Pre-denying-retweets (true): -0.30235270317617}
% \todo{Post-denying-retweets (true): -0.3162853473825}

While the evidence from retweets suggests that social media users are not very good at distinguishing between true and false stories when retweeting rumourous tweets, the actual tweet responses to rumourous tweets may reflect a different picture. In the next section we investigate the level of support and denial observed in the replies to rumourous tweets. %to understand if users responding to rumours do better in dealing with rumours when it comes to responding to them.

\subsection*{Rumour Support and Denial}

The crowdsourced annotations, which manually categorise each of the source and response tweets according to the type of support expressed with respect to the rumour, enable us to analyse the performance of social media users in terms of support and denial of rumours. Specifically, our annotation scheme categorised each tweet as supporting a rumour, denying it, appealing for more information, or making a comment. For a simplified analysis of rumour support and denial, we omit comments, which do not contribute to resolving the veracity of a rumour. We consider supporting tweets on the one hand and, on the other, we combine denying tweets and appeals for more information into denials. Having these two types of response tweets, i.e., supports and denials, we define the support ratio as the ratio of supporting tweets over those that are either supporting or denying a rumour, so that we can normalise the support and make it comparable across rumours and events. The equation that determines the support ratio for a rumour is defined in Equation \ref{eq:support-ratio}.

\begin{equation}
 support ratio = \frac{\# support}{\# support + \# denials}
 \label{eq:support-ratio}
\end{equation}

\begin{figure}[ht]
 \begin{adjustwidth}{-0.8in}{0in}
  \centering
  \includegraphics[width=1.3\textwidth]{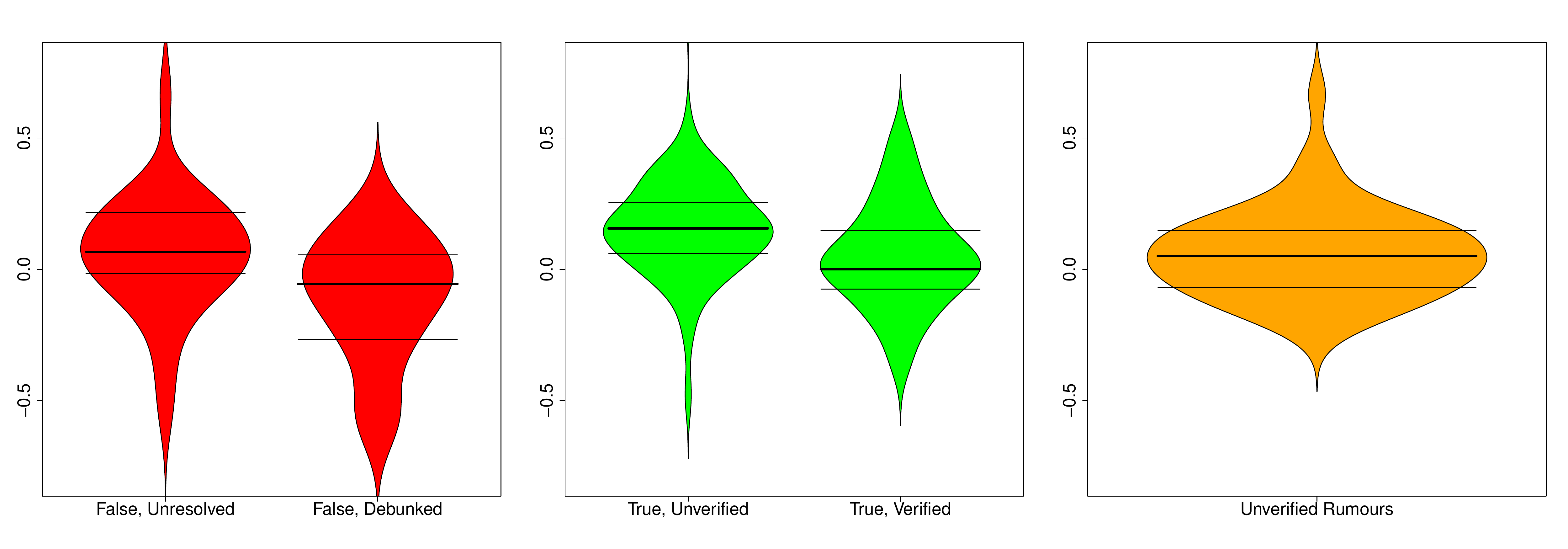}
  \caption{Distribution of support ratios before and after resolving tweets for true and false rumours, as well as for rumours that remain unverified. Horizontal lines represent 25, 50, and 75 percentiles.}
  \label{fig:support-boxplots}
 \end{adjustwidth}
\end{figure}

We look at the support ratio observed for both true and false rumours, distinguishing between ratios before and after the resolving tweet is posted, so that we can also explore the effect of the resolving tweet on public opinion. Figure \ref{fig:support-boxplots} shows support ratios before and after resolving tweets for rumours that turn out to be true or false.

We observe that true rumours tend to have a slightly higher ratio of support (median = 0.156) than false rumours (median = 0.067) before the resolving tweet appears, which is statistically significant ($W = 1099.5$, $p = 0.02014$). However, the fact that both median values are positive and the difference between the two populations in this case is relatively small, suggests that the aggregation of collective intelligence may not be sufficient to determine the veracity of a rumour, at least not before the resolving tweet is posted. While rumours remain unverified, the overall tendency is to  support them, or to assume that they are true, irrespective of their actual truth value, as shown by the prevalence of positive support ratios in these cases.

Surprisingly, the support ratio decreases after the resolving tweet for both true and false rumours, in relation to the support ratio observed before the advent of the resolving tweet. Furthermore, there is a larger difference in terms of the support ratio between true and false rumours after the advent of the resolving tweet, with a median of 0.02083 for true rumours, and a median of -0.0723 for false rumours, although without sufficient statistical significance ($W = 220$, $p = 0.07854$).

The support ratio for unverified stories is very similar to the support ratio observed for false and true rumours prior to their resolution. The support ratio for unverified stories tends to be low but slightly in favour of supporting them (median = 0.04348). This confirms the overall tendency to support unverified stories.

\begin{figure}[ht]
  \centering
  \includegraphics[width=0.5\textwidth]{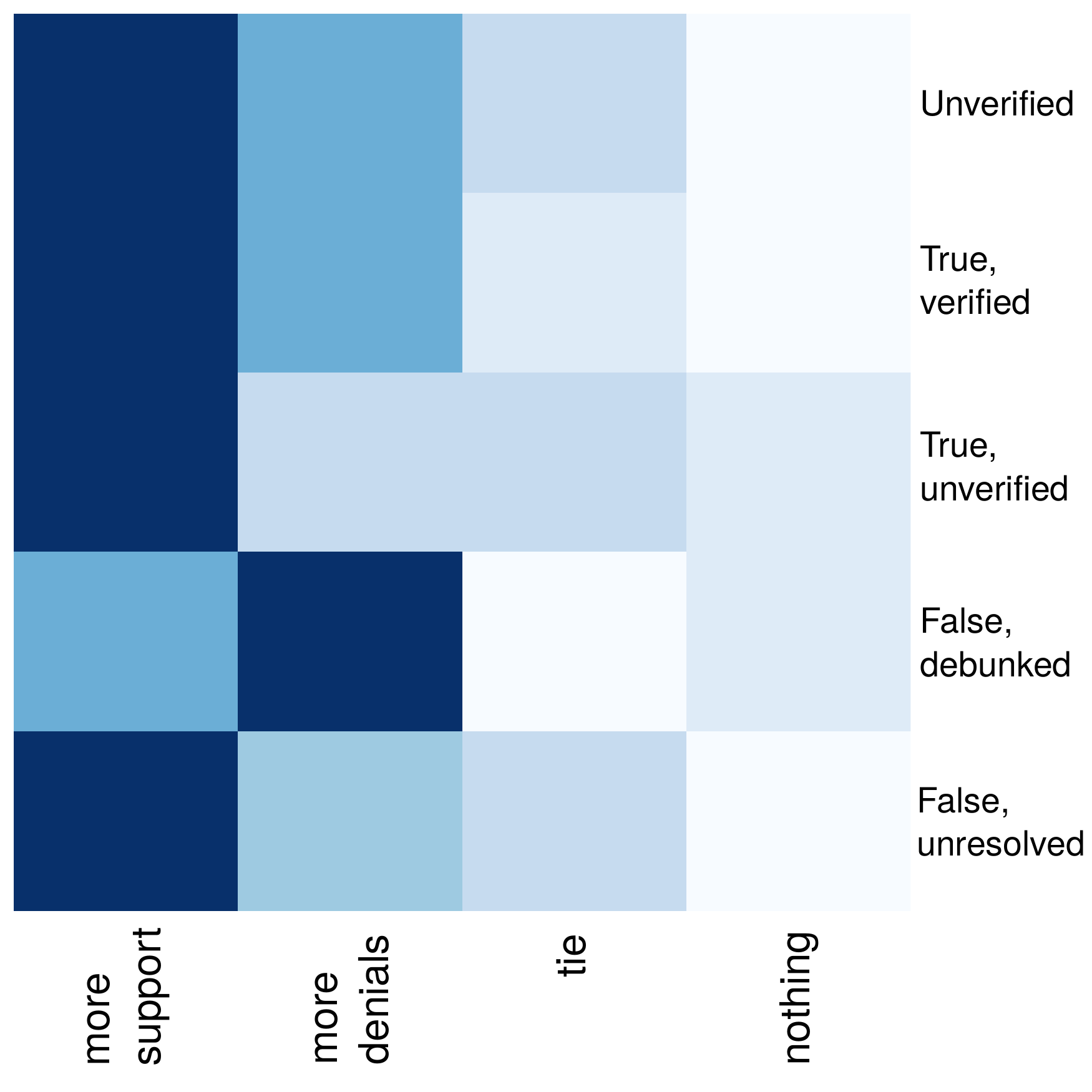}
  \caption{Support heatmap representing the number of rumours that spark (1) more support, (2) more denials, (3) the same number of each, or (4) neither.}
  \label{fig:support-heatmap}
\end{figure}

As an additional step here, we look at the number of cases in which the amount of support exceeds denial, those in which denial beats support, and those in which they have equal weight. Figure \ref{fig:support-heatmap} visualises on a heatmap the number of rumour stories where there are either more supporting or more denying tweets, there is a tie, or there is not a single tweet questioning it or backing it up. This again buttresses our previous observation that most people tend to support rumourous stories and few people question or deny them. In fact, the chart shows that the only case in which the amount of denial beats support is that of false stories, after the resolving tweet has been posted.

Apart from looking at the support ratio, we also analyse the amount of discussion that each type of rumour (true, false, pre and post resolution) produces. We define the discussion ratio as the sum of supporting and denying tweets divided by all the tweets, hence referring to the fraction of tweets that discuss the veracity of a rumour either positively or negatively (see Equation \ref{eq:discussion-ratio}).

\begin{equation}
 discussionratio = \frac{\# support + \# denials}{\# support + \# denials + \# comments}
 \label{eq:discussion-ratio}
\end{equation}

\begin{figure}[ht]
  \centering
  \includegraphics[width=0.7\textwidth]{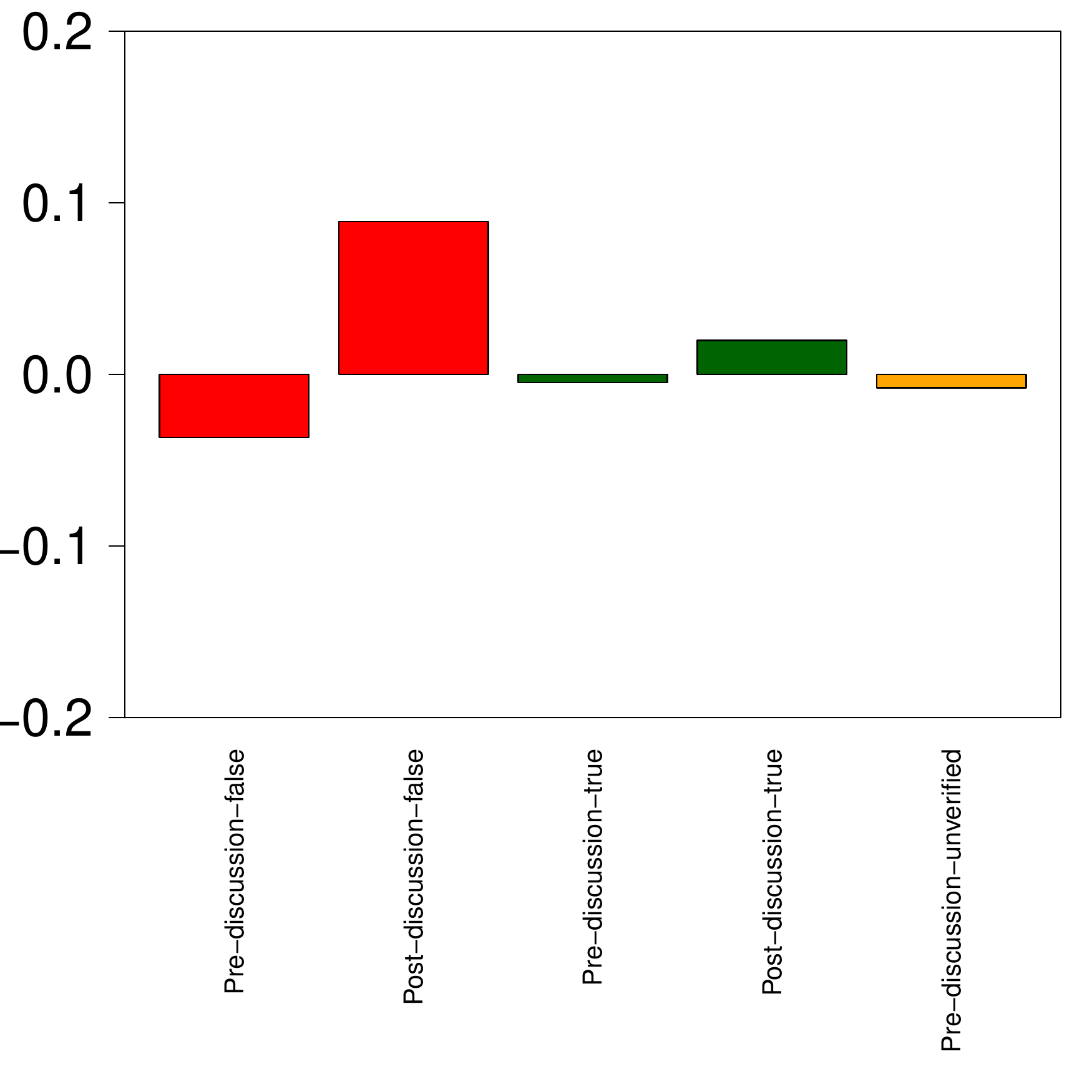}
  \caption{Z-score values for the amount of discussion generated by different rumours, representing the extent to which different rumours spark discussion below or above the overall average. Red corresponds to false rumours, green to true rumours and yellow to unverified rumours.}
  \label{fig:discussion-zscores}
\end{figure}

Figure \ref{fig:discussion-zscores} compares the amount of discussion that different types of rumourous tweets spark before and after resolving tweets. We use z-scores again to normalise the values and make them comparable to one another for rumours corresponding to different events.

Here we observe that there is more discussion after the veracity of a rumour is resolved, especially for false rumours. Unverified stories all have negative z-scores, irrespective of the stories later being proven true or false, or remaining unverified. However, positive z-scores can be observed when the resolving tweet has been made public. Interestingly, the fact that the resolving tweet has been posted produces a higher number of tweets discussing the veracity of the rumour, either positively or negatively. To further quantify the amount of support and denial produced in these situations, we look next at the support ratio.

% \todo{Pre-discussion (false): -0.036683863003092}
% \todo{Post-discussion (false): 0.089060839093678}
% \todo{Pre-discussion (true): -0.0047249337169608}
% \todo{Post-discussion (true): 0.019804923883968}
% \todo{Pre-discussion (unverified): -0.0077788595048356}

\begin{figure}[ht]
  \centering
  \includegraphics[width=0.7\textwidth]{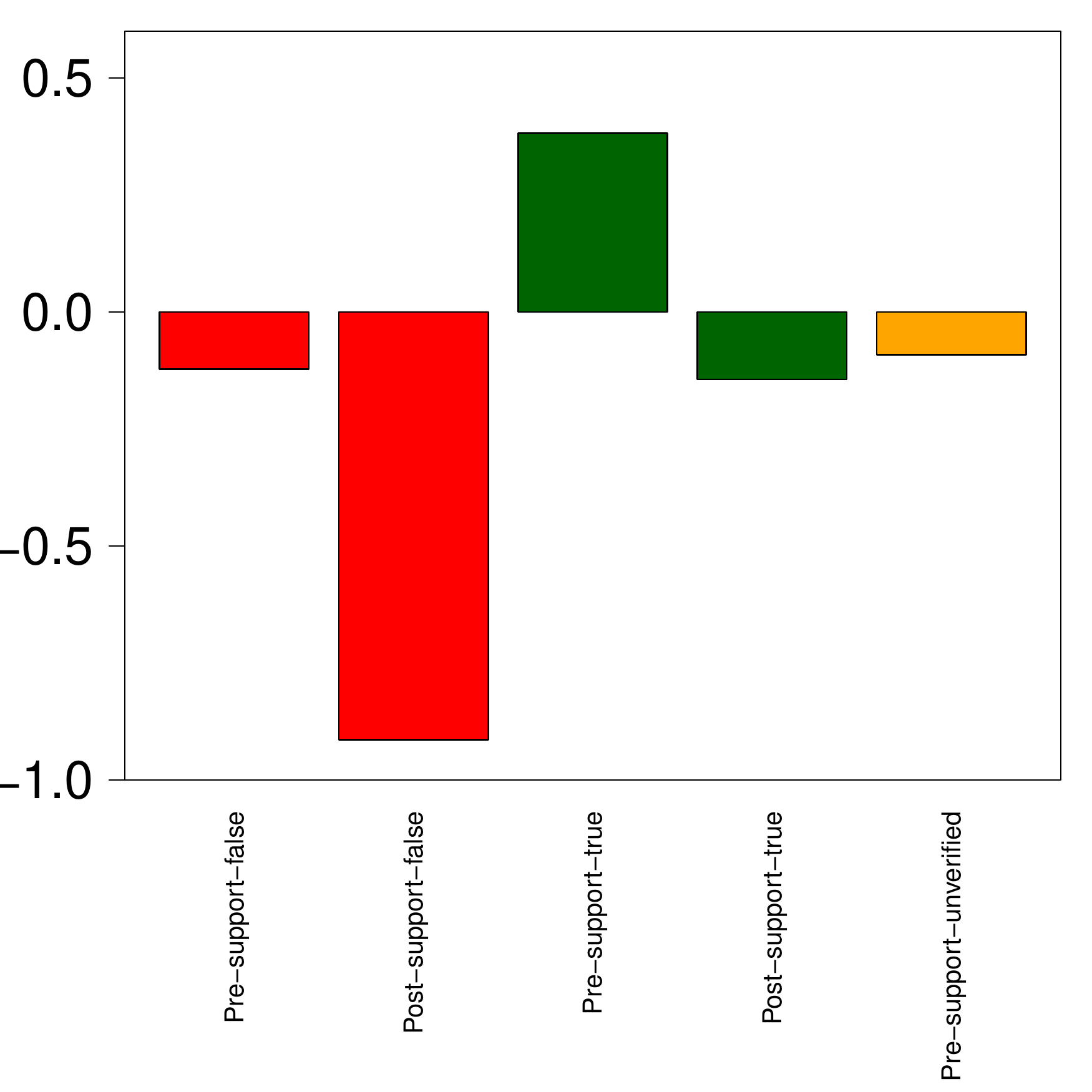}
  \caption{Z-score values for support ratios observed in different rumours, representing the extent to which different rumours get support below or above the overall average.}
  \label{fig:support-zscores}
\end{figure}

Figure \ref{fig:support-zscores} shows the support ratio that different types of tweets (true, false, pre and post resolution) receive. The large increase of discussion for false stories once veracity is resolved, as observed above, turns into denying messages. This suggests that users are good at denying already debunked rumours, but are not so good at handling unresolved rumours. When it comes to true rumours, users do quite well in determining that an unverified rumour is true. However, and surprisingly, the slight increase in discussion after a true rumour is verified turns into a higher degree of denial. We believe that this is due to the fact that supporting messages are no longer necessary, but messages from users remaining skeptical feature more prominently in the discussions.

% \todo{Pre-support (false): -0.12162610641712}
% \todo{Post-support (false): -0.91361409909308}\todo{Explain how certainty ratio is computed.}
% \todo{Pre-support (true): 0.38195922485017}
% \todo{Post-support (true): -0.14419937319167}
% \todo{Post-support (unverified): -0.091592156666563}

\subsection*{Rumour Certainty}
Certainty, as we annotated it in the context of rumours, measures the degree of confidence expressed by the author of a tweet. We look at the degree of certainty expressed by users in the context of rumours of different veracity statuses to check how it changes and whether it increases as rumours get resolved. To compute the certainty ratio observed in each of the rumourous stories, we simplify the annotations of certainty and distinguish two groups: (1) certain, which includes annotations of tweets belonging to a rumour as `certain', and (2) not-entirely-certain, which includes annotations of tweets as `somewhat certain' and `uncertain'. Given these two values, the certainty ratio is computed as:

\begin{equation}
 certainty ratio = \frac{\# certain}{\# certain + \# \text{\textit{not-entirely-certain}}}
 \label{eq:certainty-ratio}
\end{equation}

\begin{figure}[ht]
 \begin{adjustwidth}{-0.8in}{0in}
  \centering
  \includegraphics[width=1.3\textwidth]{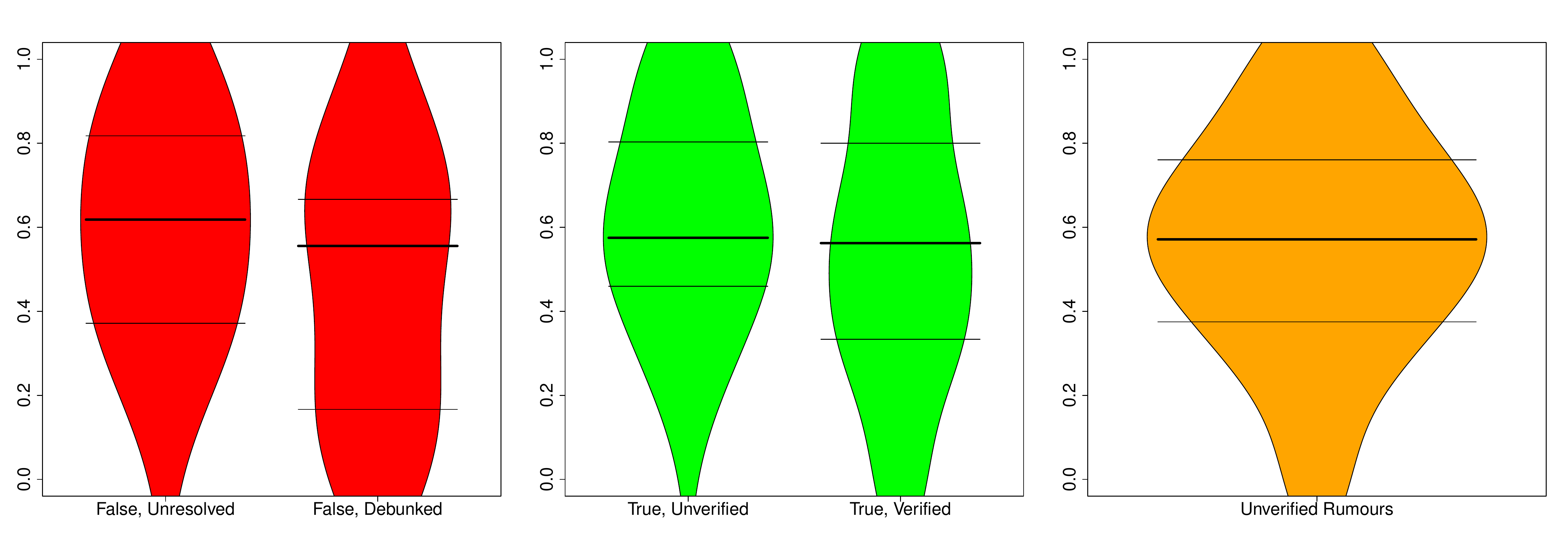}
  \caption{Distribution of certainty ratios before and after resolving tweets for true and false rumours, as well as for rumours that remain unverified. Horizontal lines represent 25, 50, and 75 percentiles.}
  \label{fig:certainty-boxplots}
 \end{adjustwidth}
\end{figure}

Figure \ref{fig:certainty-boxplots} shows the distributions of certainty ratios by veracity status (true, false, unverified). Certainty appears to be stable before and after resolving tweets for true rumours; with a median of 0.575 before being resolved, and a median of 0.563 afterwards, there is no significant difference in the degree of certainty after a rumour has been verified ($W = 848$, $p = 0.345$). However, we do observe a larger difference in the case of false rumours, where the median is 0.618 before resolution, and 0.556 thereafter; however, this difference cannot be considered to be statistically significant ($W = 224.5$, $p = 0.2384$). In the rest of the cases, i.e., unverified rumours, the certainty level (median = 0.571) is very similar to true rumours eventually verified as true.

\subsection*{Rumour Evidentiality}

Evidentiality in the context of a rumour determines the type of evidence (if any) provided by an author of a tweet. We look next at the extent to which users provide evidence with their statements for different veracity statuses. To compute the evidentiality ratio in rumourous stories, we simplify the annotations of evidence for individual tweets within a rumourous thread as follows: (1) `evidence', which applies to a tweet that provides any kind of evidence, and (2) `no-evidence', for tweets that have no evidence attached. Given these two values, we compute the evidentiality ratio as:

\begin{equation}
 evidentiality ratio = \frac{\# evidence}{\# evidence + \# \text{\textit{no-evidence}}}
 \label{eq:evidentiality-ratio}
\end{equation}

\begin{figure}[ht]
 \begin{adjustwidth}{-0.8in}{0in}
  \centering
  \includegraphics[width=1.3\textwidth]{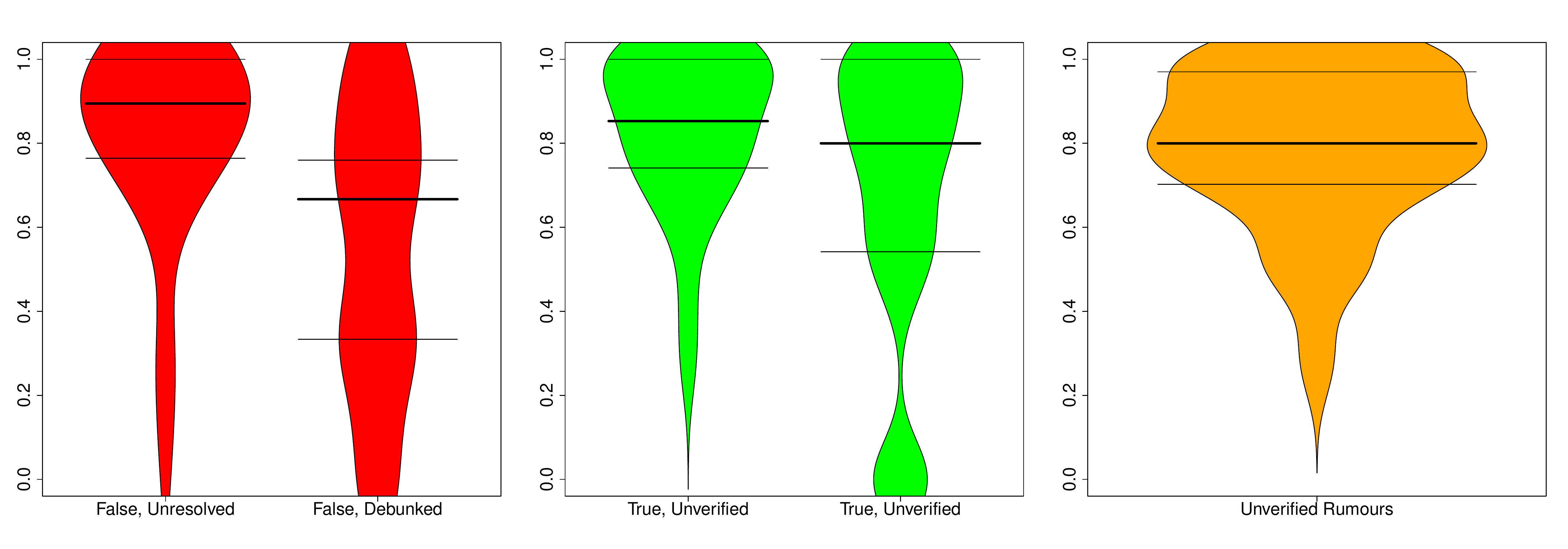}
  \caption{Distribution of evidentiality ratios before and after resolving tweets for true and false rumours, as well as for rumours that remain unverified. Horizontal lines represent 25, 50, and 75 percentiles.}
  \label{fig:evidentiality-boxplots}
 \end{adjustwidth}
\end{figure}

Figure \ref{fig:evidentiality-boxplots} shows the distributions of evidentiality ratios by veracity status. The values for evidentiality associated with rumours that are as yet unverified is very similar, irrespective of whether the rumour is proven to be false, true, or remains unverified. However, there is a noticeable drop of evidentiality after a rumour is resolved. This is especially noticeable for false rumours. Interestingly, most tweets provide supporting evidence when a false rumour is still unverified (median = 0.8944), but this number drops dramatically once the rumour is proven false (median = 0.667); this difference is statistically significant ($W = 269.5$, $p = 0.01382$). The drop in evidentiality after rumour resolution is smaller for true rumours, whose median value is 0.853, while unverified; this drops slightly to 0.8 after the advent of the resolving tweet; however, this difference cannot be deemed statistically significant ($W = 932$, $p = 0.07333$). The significant drop in evidentiality associated with false rumours once they are debunked suggests a shift in the focus of the discussion from trying to disprove a rumour.

\subsection*{Analysis of Users}

Next, we examine the role that different users play in the diffusion and support of rumours. In our rumour datasets, a total of 56,099 different users participated: 208 posted rumourous source tweets, 3,129 responded to the rumours, and 53,469 retweeted the rumours. We rank users by the amount of followers they have, which, as we argue below, can be used to provide a measure of the reputation of the user. However, the number of followers might somehow be rigged, e.g., by users who simply follow many others to attract more followers. To control for this effect, we define the follow ratio as the logarithmically scaled ratio of followers over followees (see Equation \ref{eq:follow-ratio}).

\begin{equation}
 \log_{10} \frac{\# followers}{\# following}
 \label{eq:follow-ratio}
\end{equation}

Hence, the larger the number of followers a user has in relation to the number of users they are following, the larger the follow ratio will be.

To assess the validity of the follow ratio as a reputation measure, we took a closer look at users with a high follow ratio. Within the possible follow ratio values, which ranges from -1 to 7 in our dataset, we looked at the 42 users with a follow ratio of 4 or higher. Among these 42 highly reputable users, 38 are news organisations, % 20Minutes, ABC, AJENews, AJEnglish, AlArabiya_Eng, BBCBreaking, BBCWorld, BILD, BILD_News, Breaking911, BreakingNews, CBCAlerts, CBSNews, ChannelNewsAsia, cnnbrk, cnni, CP24, FoxNews, FRANCE24, globeandmail, guardian, guardiannews, mashable, Reuters, RT_com, SkyNews, SkyNewsBreak, Slate, SPIEGELONLINE, SPIEGEL_Top, staronline, tagesschau, Telegraph, TIME, USATODAY, welt, WSJ, zeitonline
2 are 'parody accounts' (e.g. accounts self-evidently set up to make fun of or 'spoof' real people or organisations), % thelittleidiot, @Der\_Oberlehrer
1 is an activist, % daxshepard1
and 1 is an actor. % MichaelSkolnik
Hence, the majority of users with high follow ratios in the context of newsworthy events are news organisations, which validates the follow ratio as a reputation measure for our purposes.

\begin{figure}[ht]
 \begin{adjustwidth}{-0.8in}{0in}
  \centering
  \includegraphics[width=1.3\textwidth]{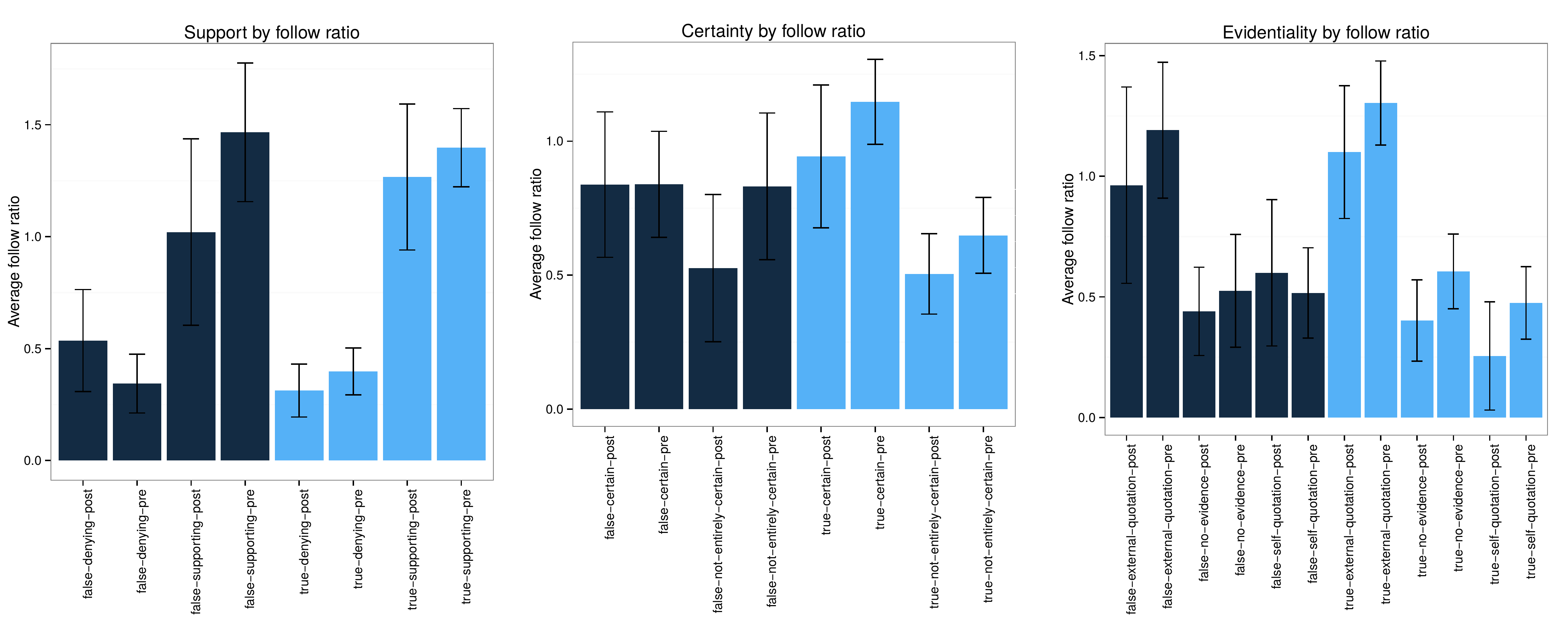}
  \caption{Average follow ratios for users who express different types of support, certainty and evidentiality, along with error bars that represent the 95\% confidence interval.}
  \label{fig:users-by-follow-ratio}
 \end{adjustwidth}
\end{figure}

Figure \ref{fig:users-by-follow-ratio} shows the statistics for the average follow ratio of users who express support, certainty and evidentiality for rumours at different veracity status levels. Here we observe that users who support rumours, irrespective of whether they are ultimately true or false, tend to have a substantially higher follow ratio compared to those users who challenge them. On the contrary, users who deny rumours seem to have a lower follow ratio, irrespective of whether the story is ultimately true or false. This divergence can actually be observed both while rumours are still unverified, as well as later, when their veracity has been resolved. When we look at the relationship between follow ratio and certainty, we see that users who tend to be certain are those with a higher follow ratio, whereas users with lower follow ratios tend to be uncertain. This seems to be the case for both true and false rumours, although uncertainty is expressed by users with a high follow ratio in the case of false rumours prior to their resolution. Looking at the types of evidence that users of different levels of reputation attach to their tweets, we can see some interesting patterns that differentiate users with low and high follow ratios. In fact, users with high follow ratios tend to quote external pieces of evidence to back up their own posts, such as a link to a news story, while users with a low follow ratio tend to either not provide any evidence at all or rely on self-assertion. 

Given that the majority of users in our sample with a high follow ratio are news organisations, these results do not seem very surprising. One the one hand, we would expect journalists to orientate to the need to display compliance with professional values in reporting stories, which is consistent with quoting sources, even if these sources themselves may not be entirely trustworthy \cite {cotter2010news}. On the other hand, the pressure to produce something in the face of an over-abundance of information and the presence of numerous near-identical stories from competing news organisations means that verification of a rumour may take second place to the need to publish \cite{boczkowski2010news}. 

Apart from the user follow ratio, we also considered other factors to distinguish between users, such as user age, whether or not they are verified users, or the number of times they tweet, but found no significant differences.

\section*{Discussion and Conclusions}

The methodology we have developed for collecting, identifying, and annotating rumour threads has enabled us to analyse a number of important aspects of how people react to rumours using measures of diffusion and in the context of responses that either support or deny the veracity of the rumours. The application of this methodology to 330 rumour threads associated with 9 different newsworthy events and 4,842 tweets has helped us discover behavioural patterns that can be observed in the context of rumours. To the best of our knowledge, this is the first work to perform such a comprehensive study on a diverse set of rumours. We consider conversational aspects of rumours and also differentiate between two parts of the rumour cycle, i.e., prior to the resolution of its veracity status and after a rumour has been debunked or verified. To summarise what has come out of the study results:

\textbf{True rumours tend to be resolved faster than false rumours.} Our study shows a significant difference in terms of the time it takes before true and false rumours are resolved respectively. While most true rumours are corroborated within 2 hours of the source tweet that introduces them, false rumours tend to take about 14 hours before they are debunked. This observation aligns with common sense and buttresses earlier conjectures according to which proving that a fact is not accurate is far more difficult than proving it is true \cite{saunders1984mythic}; providing counter-evidence that disproves a statement is indeed more challenging.

\textbf{Rumours in their unverified stage produce a distinctive burst in the number of retweets within the first few minutes, substantially more than rumours proven true or false.} When we looked at the diffusion of rumours in the form of retweets, we observed a clear tendency among users to share rumours that are still in the first phase of their life cycle, prior to them being resolved. Our analysis shows that tweets reporting nverified rumours spark a sudden burst of retweets within the very first minutes, showing that users tend to share early, unverified reports rather than later confirmations or debunks. This might seem somewhat surprising at first, as one would expect people to be more interested in the final resolution of a rumour. However, this observation aligns with the nature of rumours, which by definition produce increased levels of anxiety, increasing, in turn, the likelihood of passing on information to others. Our analysis suggests that interest in spreading a story decreases after its veracity value is resolved, especially when it is proven false. This finding also aligns with investigations of the reportability of newsworthy information in the conversation analytic literature \cite{sacks1992lectures}.

\textbf{The prevalent tendency of users is to support every unverified rumour.} Interesting insights into rumours are also obtained from our analysis of discussion in the form of supporting and denying tweets. When we considered both support and denial tweets together, which we referred to as discussion, we observed that the amount of discussion increases substantially once a rumour is resolved, irrespective of whether it is corroborated or debunked. When distinguishing between support and denials, we observe that it is denial which increases after rumour resolution, and support rather drops off as a rumour is resolved, irrespective of whether the story turns out to be true or false. It is surprising that an increase in messages denying a rumour not only occurs with rumours that have been debunked, but also in the case of those confirmed to be true. This suggests that those who trust the confirmation no longer feel the need to insist, while a number of skeptical users emerge rebutting the resolution of the rumour. It is worth noting that the increase in messages denying a rumour does not usually imply that they numerically exceed supporting messages. In fact, the only case in which denying messages outnumber supporting messages is when a false rumour has been debunked; this does not occur with false rumours while they are still unverified. Overall, the prevalent tendency of users is to support every unverified rumour, potentially because of the lack of counter-evidence.

\textbf{While the level of certainty does not change over the course of the rumour lifecycles, users provide evidence in their tweets when rumours are yet to be resolved, but less so after resolution.} Apart from looking at whether tweets were supporting or denying rumours, we also looked at certainty -- the level of confidence expressed by the author of a tweet within the conversation thread -- and evidentiality -- whether there is evidence linked to tweets within a rumourous conversation thread. We found that there is not much of a difference in terms of certainty for different veracity statuses of rumours (true, false, unverified), other than debunked rumours being associated with a slightly lower level of certainty. This is potentially due to posts by skeptics, but the finding was not statistically significant. In terms of evidentiality, there is a significant drop in its presence in tweets after a false rumour has been debunked, in comparison to tweets posted at an earlier stage, when the rumour was still unverified.

\textbf{Highly reputable users such as news organisations tend to support rumours, irrespective of them being eventually confirmed or debunked, tweet with certainty and provide evidence within their tweets.} Finally, we have also looked at different characteristics of users to find out whether different types of users react differently to rumours. Among the characteristics studied, we found significant differences when separating users by follow ratio -- i.e., the scaled ratio of the number of a user's followers to the number of accounts the user follows. Our analysis shows that users with high follow ratios are more likely to: (1) support any rumour, irrespective of its truth value; (2) be certain about their statements and (3) attach evidence to their tweets by quoting an external source. On the other hand, users with low follow ratios are more likely to: (1) deny rumours, irrespective of their actual truth value; (2) be rather uncertain about their statements and (3) either provide no evidence in their tweets, or provide evidence on the basis of their own experience, opinions or observations. 

In our context of newsworthy events the majority of users with high follow ratios are news organisations, which we might reasonably assume are at pains to act in ways that are consistent with professional standards when it comes to news reporting and so justify their readers' trust. Hence, we observe that news organisations tend to make well-grounded postings; they usually do so with certainty and often quote an external source as evidence. We also observe that despite their evident endeavour to make well-grounded statements, news organisations may nevertheless post rumours that are eventually proven false. This finding is not necessarily inconsistent with acting to maintain professional reputation, but reflects how this has to be balanced against the pressures journalists may face to publish, even when stories and sources are unverified. Hence, quoting a source, whether verified or not, represents a minimum threshold for journalists to be seen as adhering to professional standards and so be judged to be trustworthy. This finding, in fact, buttresses the observations on digital media ethics by Ward \cite{ward_digital}, who posited that values in on-line journalism have shifted from the tradition of `accuracy, pre-publication verification, balance, impartiality, and gate-keeping' to `immediacy, transparency, partiality and post-publication correction'. This has brought about `a tension between traditional journalism and on-line journalism'. 

%While this analysis of users has brought to light important aspects regarding news organisations' publication practices in the context of rumours, a more detailed analysis of users as well as looking at other users beyond news organisations would make a solid contribution extending this research, which is not within the scope of our work.

%\todo{What our sample does not enable us to test is whether other kinds of users with high follow ratios display a similar obligation to substantiate their postings with evidence: i.e., does having a high follow ratio tend to make social media users more careful in what they post and thus, in some sense, more trustworthy?}

\textbf{Future work.} Our study sheds new light on human reactions to rumours and provides insights towards better understanding of how to control the spread of rumours. One of our next objectives is to use the datasets we have collected to develop machine learning techniques for automatically predicting rumour veracity at different time points and using conversational patterns to help establish the veracity of rumours. In this way, we believe it will be possible to assist journalists to check their sources more quickly and thoroughly, and thus enable them to play a more effective role in mitigating the spread and impact of false rumours. We would also like to verify whether our observation of the tendency for users with high follow ratios to behave more responsibly can generalise from news organisations to other types of users with high follow ratios.

% Do NOT remove this, even if you are not including acknowledgments.
\section*{Acknowledgments}

This work has been supported by the PHEME FP7 project (grant No. 611233).

% \nolinenumbers

\bibliographystyle{apalike}
\bibliography{arxiv-rumours}

\end{document}